\documentclass[sigconf, nonacm, pdfa]{acmart}

\usepackage[a-2b]{pdfx}
\usepackage{hyperref}
\usepackage{balance}
\usepackage[compress]{cleveref}
\usepackage{caption}
\usepackage{subcaption}
\usepackage{xspace}
\usepackage{amsmath}
\usepackage{array, caption, booktabs}
\usepackage[linesnumbered,ruled,noend]{algorithm2e}
\usepackage{setspace}
\usepackage[title]{appendix}
\usepackage{bbm}
\usepackage{multirow}
\usepackage{subcaption}
\usepackage{listings}
\usepackage{color}
\usepackage{enumitem}
\usepackage{graphicx}
\usepackage{makecell}
\usepackage{hhline}

\definecolor{dkgreen}{rgb}{0,0.6,0}
\definecolor{gray}{rgb}{0.5,0.5,0.5}
\definecolor{mauve}{rgb}{0.58,0,0.82}
\usepackage{lipsum}

\crefname{section}{Section}{Sections}
\Crefname{section}{Section}{Sections}
\crefname{figure}{Figure}{Figures}
\Crefname{figure}{Figure}{Figures}
\crefname{subfigure}{Figure}{Figures}
\Crefname{subfigure}{Figure}{Figures}
\crefrangelabelformat{subfigure}{#3#1#4--#5(\crefstripprefix{#1}{#2}#6}

\newcommand{\system}{EQUI-VOCAL\xspace}

\setcopyright{none}
\renewcommand\footnotetextcopyrightpermission[1]{}
\AtBeginDocument{%
  \providecommand\BibTeX{{%
    \normalfont B\kern-0.5em{\scshape i\kern-0.25em b}\kern-0.8em\TeX}}}

\begin{document}

\title{EQUI-VOCAL: Synthesizing Queries for Compositional Video Events from Limited User Interactions}
\subtitle{Technical Report\vspace{-0.5em}}

\author{Enhao Zhang}
\affiliation{%
  \institution{University of Washington}
  \city{}
  \country{}
}
\email{enhaoz@cs.washington.edu}

\author{Maureen Daum}
\affiliation{%
  \institution{University of Washington}
  \city{}
  \country{}
}
\email{mdaum@cs.washington.edu}

\author{Dong He}
\affiliation{%
  \institution{University of Washington}
  \city{}
  \country{}
}
\email{donghe@cs.washington.edu}

\author{Brandon Haynes}
\affiliation{%
  \institution{Microsoft Gray Systems Lab}
  \city{}
  \country{}
}
\email{brandon.haynes@microsoft.com}

\author{Ranjay Krishna}
\affiliation{%
  \institution{University of Washington}
  \city{}
  \country{}
}
\email{ranjay@cs.washington.edu}

\author{Magdalena Balazinska}
\affiliation{%
  \institution{University of Washington}
  \city{}
  \country{}
}
\email{magda@cs.washington.edu}
\renewcommand{\shortauthors}{}

\begin{abstract}

We introduce \system: a new system that automatically synthesizes queries over videos from limited user interactions. The user only provides a handful of positive and negative examples of what they are looking for. \system utilizes these initial examples and additional ones collected through active learning to efficiently synthesize complex user queries. Our approach enables users to find events without database expertise, with limited labeling effort, and without declarative specifications or sketches. Core to \system's design is the use of spatio-temporal scene graphs in its data model and query language and a novel query synthesis approach that works on large and noisy video data. Our system outperforms two baseline systems---in terms of F1 score, synthesis time, and robustness to noise---and can flexibly synthesize complex queries that the baselines do not support.

\end{abstract}

\maketitle

\begin{sloppypar}
\newcommand{\systemTeaserFigure}{
    \begin{figure}[t!]
        \centering
        \includegraphics[width=\columnwidth]{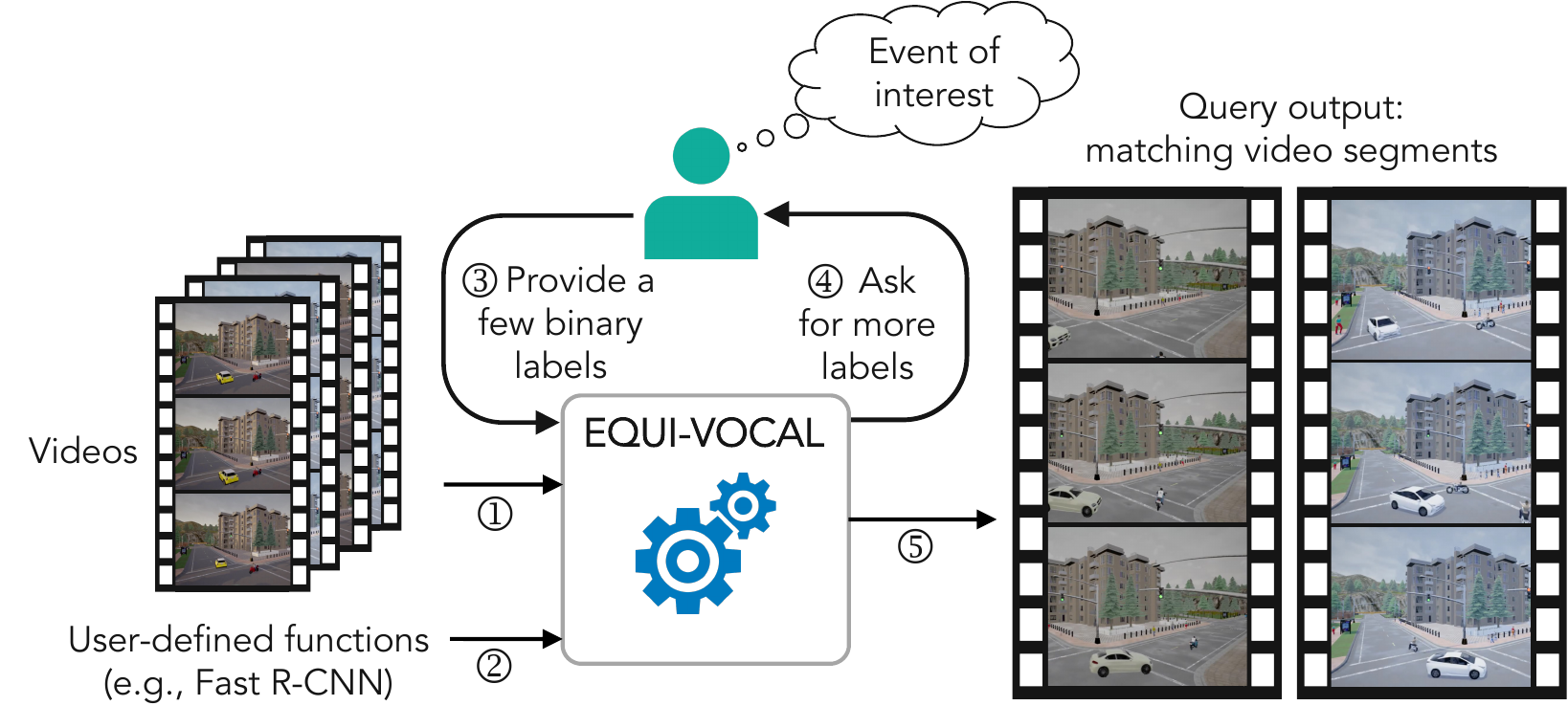}
        \vspace{-2em}
        \caption{Given \textcircled{\small 1} a video dataset,
          \textcircled{\small 2} user-defined functions that extract semantic information from videos,
          and \textcircled{\small 3} a few user-provided
          labels, \system synthesizes a query to find
          instances of an event of interest. It iteratively
          \textcircled{\small 4} asks the user for more labels to
          reduce its uncertainty. Once synthesized, it \textcircled{\small 5} executes the query to return matching events on unseen videos.}
        \vspace{-0.6cm}
        \label{fig:system_teaser}
    \end{figure}
}

\newcommand{\exampleFramesFigure}{
    \begin{figure}[t!]
        \centering
        \begin{subfigure}{0.48\columnwidth}
            \includegraphics[width=\textwidth]{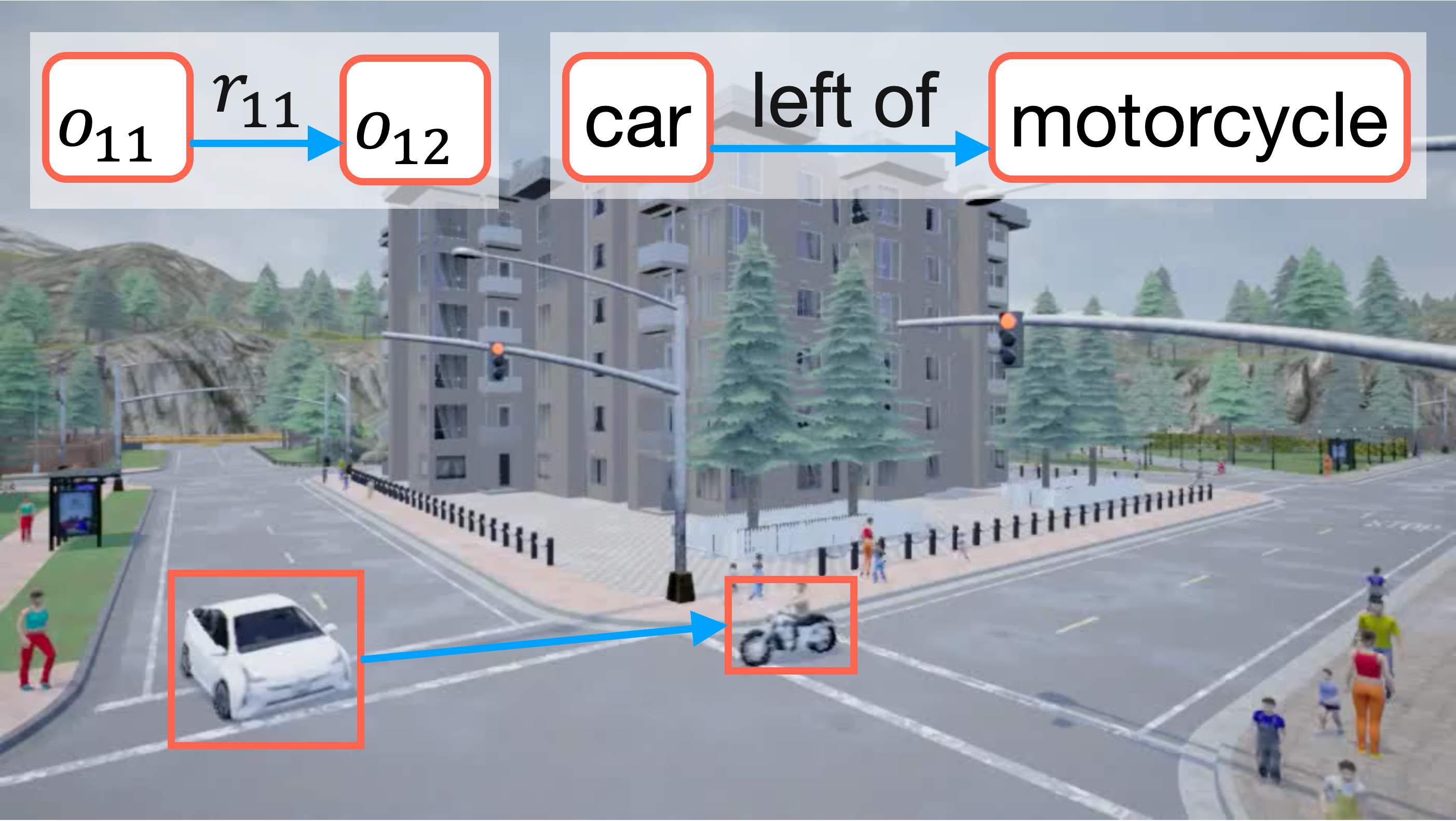}
        \end{subfigure}
        \begin{subfigure}{0.48\columnwidth}
            \includegraphics[width=\textwidth]{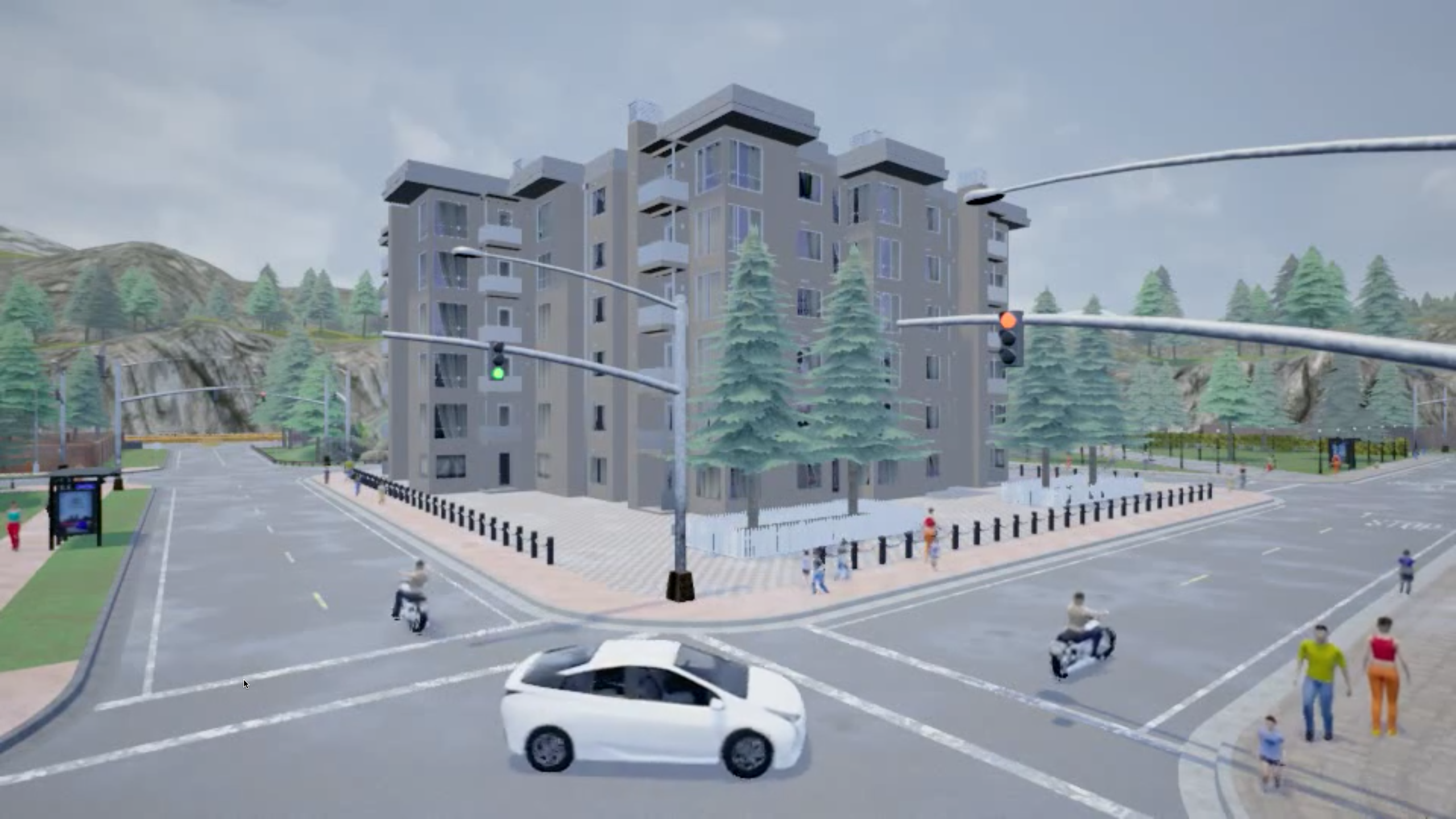}
        \end{subfigure}
        \caption{Example frames of multiple, simultaneous car-motorcycle interactions (generated using~\cite{DBLP:conf/sigmod/HaynesMBCC19}). \system represents video content as a sequence of region graphs in its data model. Each region graph models a single video frame (left figure), with nodes representing objects and edges representing relationships. A region graph is a subset of the full scene graph (not shown).}
        \label{fig:exampleFrames}
    \end{figure}
}

\newcommand{\algorithmExampleFigure}{
    \begin{figure*}[t!]
        \centering
        \includegraphics[width=\textwidth]{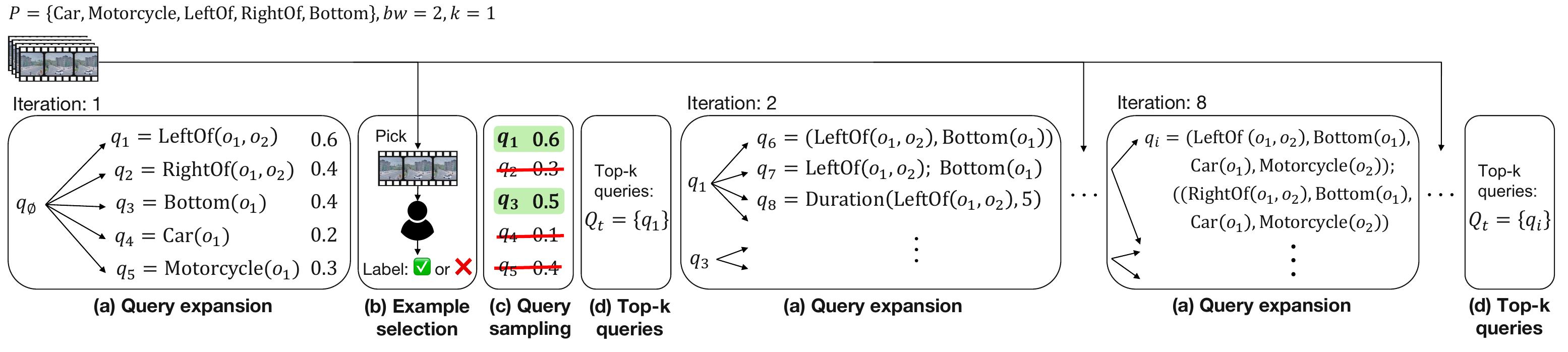}
        \caption{Running example of the query synthesis algorithm. The algorithm starts
          with $q_\emptyset$ and synthesizes queries by iteratively
          (a) expanding queries, (b) selecting new video segments for
          the user to label, (c) sampling a small set of queries for
          further expansion, and (d) updating a list of top-$k$
          queries after each iteration. The algorithm returns $Q_t=\{q_i\}$.}
        \label{fig:algorithmExample}
    \end{figure*}
}

\newcommand{\exampleExpansionFigure}{
    \begin{figure}[t!]
        \centering
        \begin{subfigure}{0.4\columnwidth}
            \includegraphics[width=\textwidth]{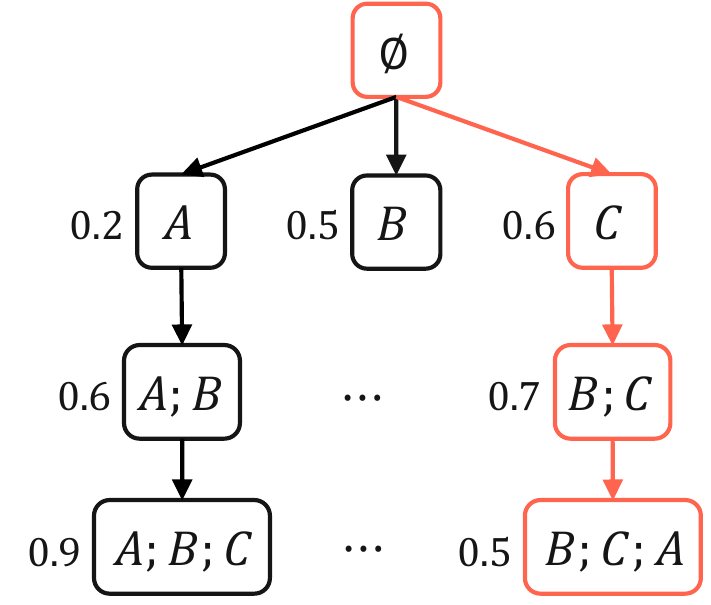}
            \caption{Restrictive expansion.}
            \label{fig:exampleExpansiona}
        \end{subfigure}
        \hspace{0.1\columnwidth}
        \begin{subfigure}{0.4\columnwidth}
            \includegraphics[width=\textwidth]{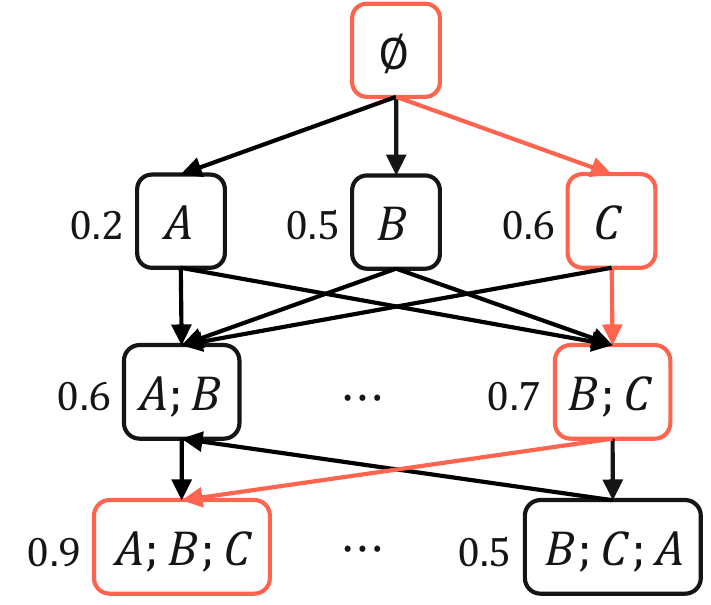}
            \caption{Relaxed expansion.}
            \label{fig:exampleExpansionb}
        \end{subfigure}
        \vspace{-1em}
        \caption{A restrictive rule constrains each query to have one construction path, while a relaxed rule allows for multiple paths. Though a relaxed rule results in a larger search space, it is more likely to find queries with high performance than a restrictive rule.}
        \label{fig:exampleExpansion}
    \end{figure}
}

\newcommand{\executionExampleFigure}{
    \begin{figure}[t!]
        \centering
        \includegraphics[width=0.9\columnwidth]{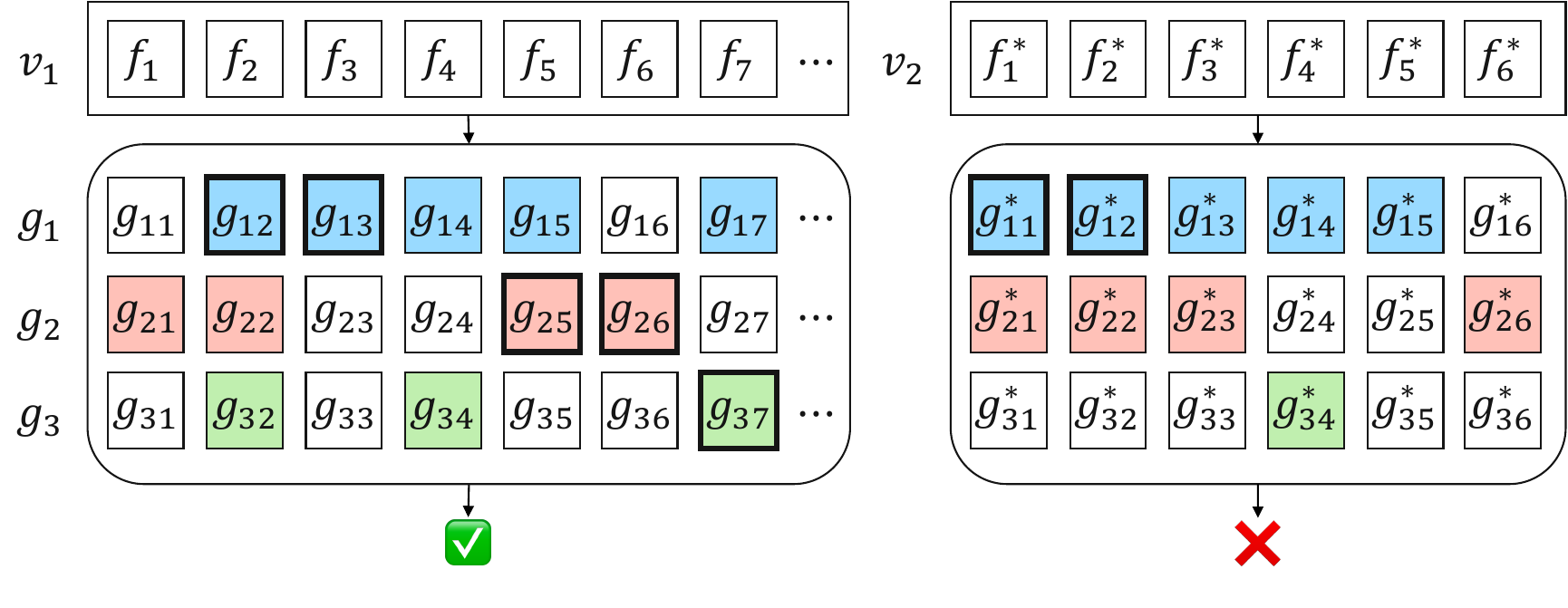}
        \vspace{-1.5em}
        \caption{Query execution example. \system reduces intermediate result sizes by computing the \textit{earliest} matching sequences for each region graph specification.}
        \label{fig:executionExample}
    \end{figure}
}

\newcommand{\withoutDurationFigure}{
    \begin{figure*}[t!]
        \centering
        \includegraphics[width=0.9\textwidth]{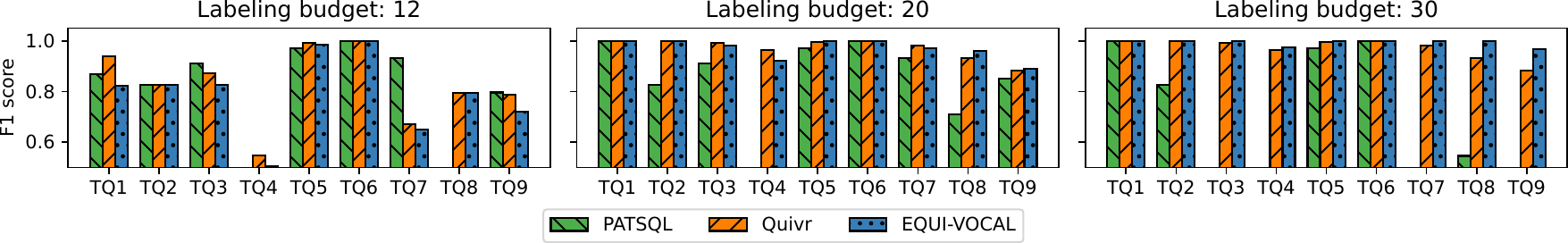}
        \caption{F1 score for queries without duration constraints on simplified tasks, under different user labeling budgets. PATSQL performs well only when the target query is simple. \system performs worse than Quivr when no additional examples are requested, but catches up and outperforms Quivr with a larger labeling budget.}
        \label{fig:withoutDuration}
    \end{figure*}
}

\newcommand{\noisyFigure}{
    \begin{figure}[t!]
        \centering
        \includegraphics[width=0.9\columnwidth]{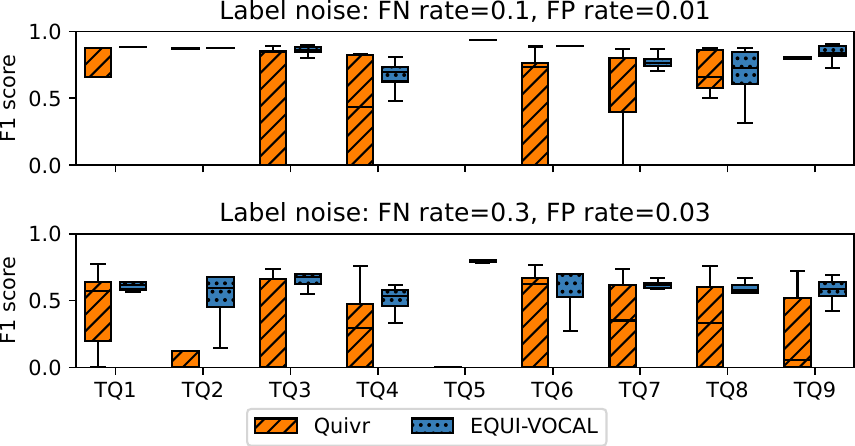}
        \caption{Impact of data noise, with a labeling budget of 20. \system constantly outperforms Quivr and is thus more robust to data noise.}
        \label{fig:noisy}
    \end{figure}
}

\newcommand{\sceneGraphQueriesFigure}{
    \begin{figure}[t!]
        \centering
        \includegraphics[width=\columnwidth]{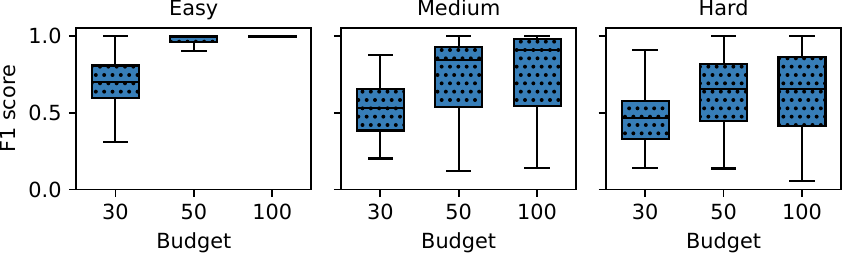}
        \caption{\system can synthesize high-quality queries within a reasonable budget, even for scene graph queries.}
        \label{fig:sceneGraphQueries}
    \end{figure}
}

\newcommand{\varycpuFigure}{
    \begin{figure}[t!]
        \centering
        \includegraphics[width=0.5\columnwidth]{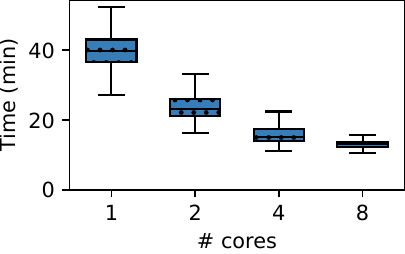}
        \caption{Vary \# cores. \system can be easily parallelized to reduce synthesis time.}
        \vspace{0.5cm}
        \label{fig:varycpu}
    \end{figure}
}

\newcommand{\warsawFullFigure}{
    \begin{figure}[t!]
        \centering
        \begin{subfigure}{\columnwidth}
            \includegraphics[width=\columnwidth]{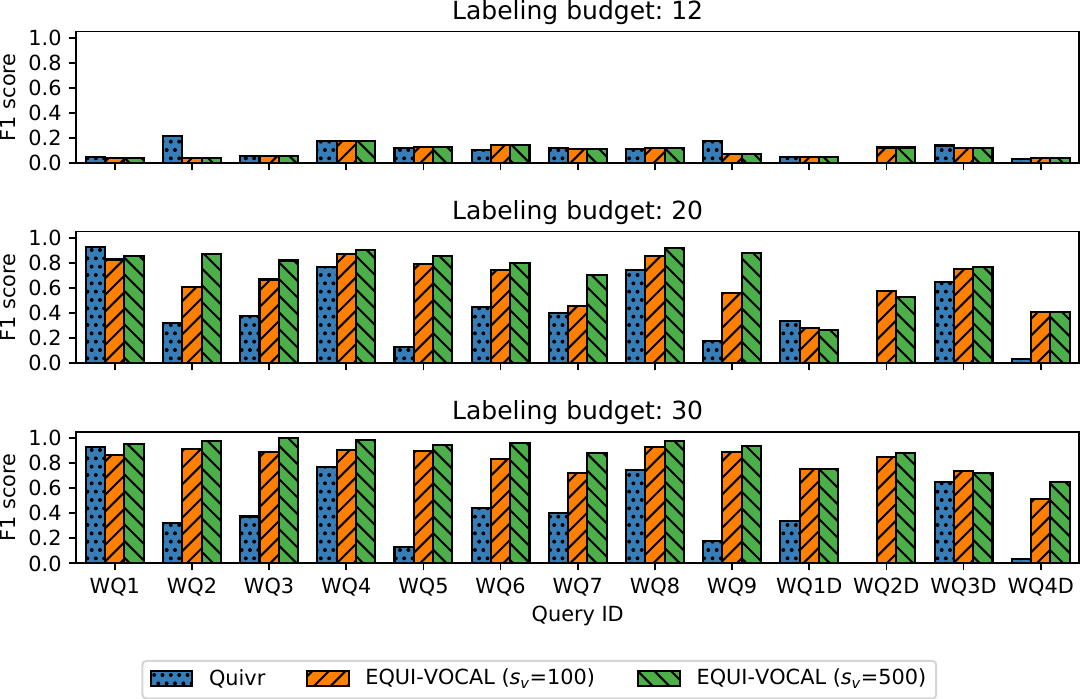}
            \caption{F1 scores.}
            \vspace{1em}
            \label{fig:warsaw_scores}
        \end{subfigure}
        \begin{subfigure}{\columnwidth}
            \includegraphics[width=\columnwidth]{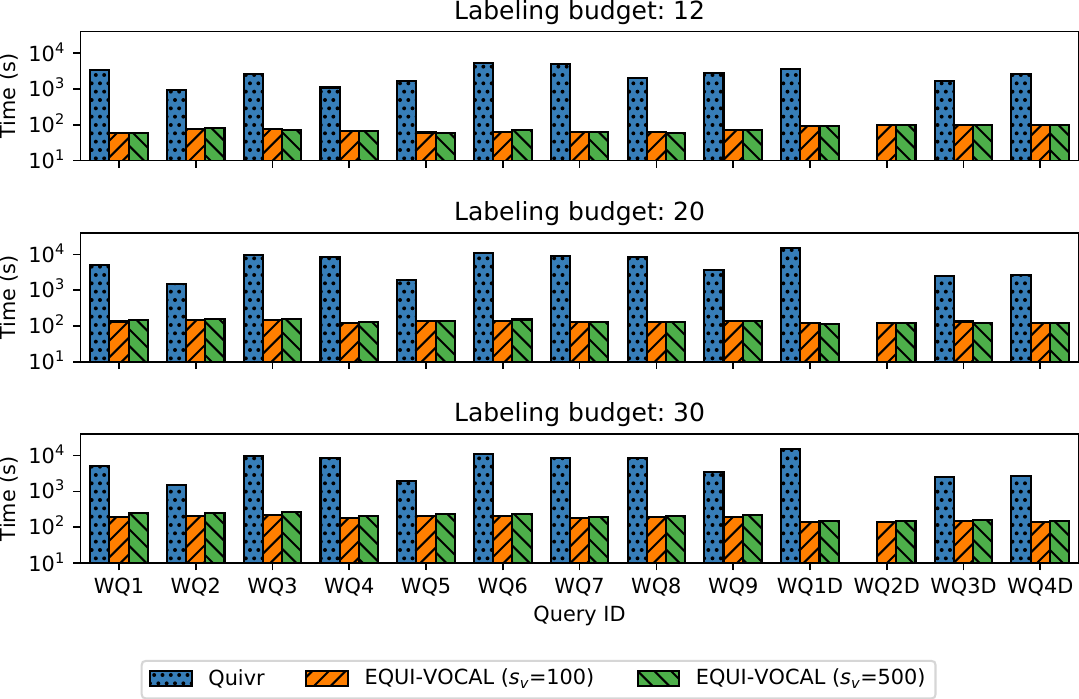}
            \caption{Query synthesis time.}
            \label{fig:warsaw_runtime}
        \end{subfigure}
        \caption{F1 scores and query synthesis time of \system and Quivr on the real-world dataset. \system can synthesize queries with high performance and is significantly faster than Quivr.}\label{fig:warsaw_full}
    \end{figure}
}

\newcommand{\userStudyFigure}{
    \begin{figure}[t!]
        \centering
        \begin{subfigure}{0.38\columnwidth}
            \includegraphics[height=2.6cm]{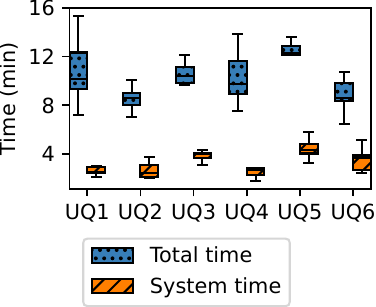}
            \caption{Task completion time.}
            \label{fig:user_study_time}
        \end{subfigure}
        \begin{subfigure}{0.58\columnwidth}
            \includegraphics[height=2.6cm]{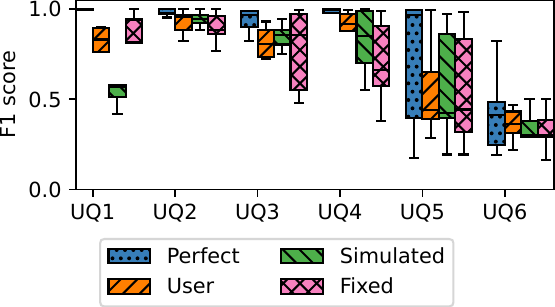}
            \caption{F1 score.}
            \label{fig:user_study_f1}
        \end{subfigure}
        \vspace{-1em}
        \caption{User study results. (a) Participants can complete the task in a reasonable time. (b) Participants can help \system synthesize UQ1-UQ4 with at least 0.8 F1 scores.}\label{fig:user_study}
    \end{figure}
}

\newcommand{\activeLearningFigure}{
    \begin{figure}[t!]
        \centering
        \includegraphics[width=\columnwidth]{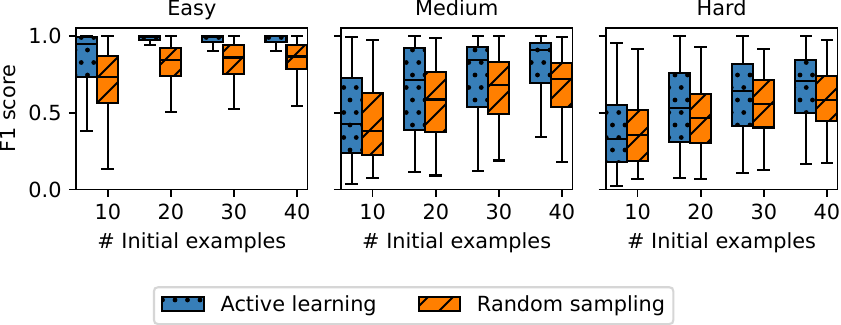}
        \vspace{-2em}
        \caption{Active learning helps \system learn better queries compared to random sampling, and the improvement is more significant when the number of initial examples is larger.}
        \label{fig:activeLearning}
    \end{figure}
}

\newcommand{\varykFigure}{
    \begin{figure}[t!]
        \centering
        \includegraphics[width=\columnwidth]{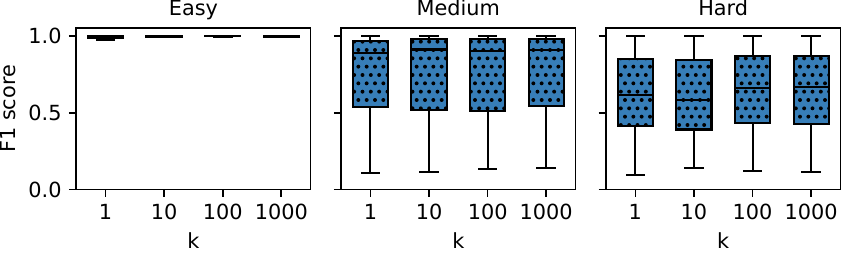}
        \vspace{-2em}
        \caption{Increasing $k$ slightly improves the performance of \system.}
        \vspace{0.2cm}
        \label{fig:varyk}
    \end{figure}
}

\newcommand{\varyInitExamplesFigure}{
    \begin{figure}[t!]
        \centering
        \includegraphics[width=\columnwidth]{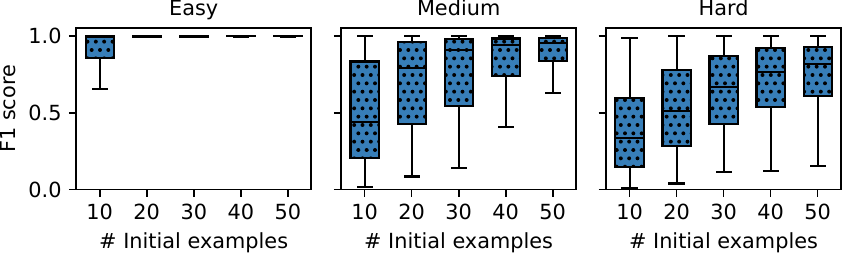}
        \vspace{-2em}
        \caption{Providing more initial examples helps \system learn better queries and avoid overfitting.}
        \vspace{-0.2cm}
        \label{fig:varyInitExamples}
    \end{figure}
}

\newcommand{\heatmapFigure}{
    \begin{figure}[t!]
        \centering
        \includegraphics[width=0.5\columnwidth]{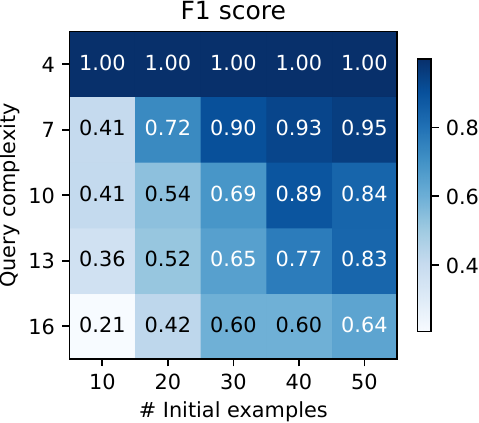}
        \caption{F1 score for various numbers of initial examples and queries with different complexity. \system obtains higher F1 scores with more initial examples and when the target queries are simpler.}
        \label{fig:heatmap}
    \end{figure}
}

\newcommand{\varybwFigure}{
    \begin{figure}[t!]
        \centering
        \includegraphics[width=\columnwidth]{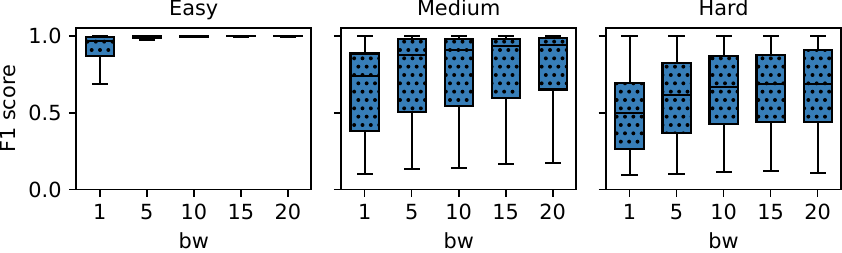}
        \caption{Increasing $bw$ improves the performance of \system.}
        \label{fig:varybw}
    \end{figure}
}

\newcommand{\varylambdaFigure}{
    \begin{figure*}[t!]
        \centering
        \includegraphics[width=0.9\textwidth]{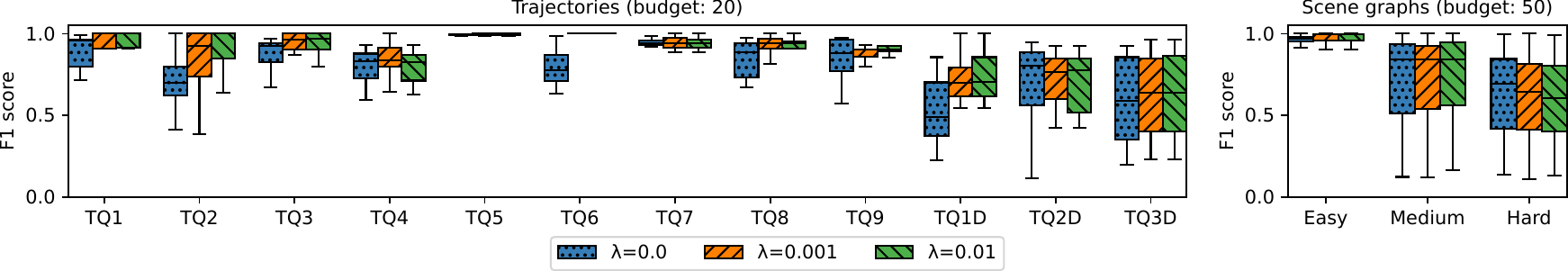}
        \caption{Regularization ($\lambda=0.001$ and $\lambda=0.01$) helps \system to learn better queries in many cases and does not harm the performance in other cases.}
        \vspace{-0.2cm}
        \label{fig:vary_lambda}
    \end{figure*}
}

\newcommand{\queryExecutionFigure}{
    \begin{figure}[t!]
        \centering
        \begin{subfigure}{0.6\columnwidth}
            \includegraphics[height=2.3cm]{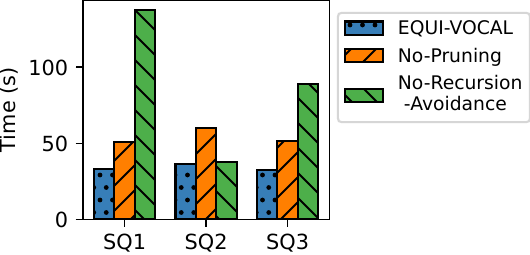}
            \caption{Single query execution time with \newline different SQL optimizations.}
            \label{fig:query_execution}
        \end{subfigure}
        \begin{subfigure}{0.35\columnwidth}
            \includegraphics[height=2.3cm]{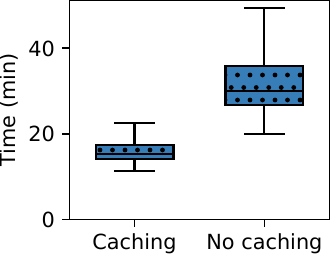}
            \caption{Synthesis time with and without caching.}
            \label{fig:caching}
        \end{subfigure}
        \caption{Impact of query execution optimizations. \system reduces query synthesis time by (a) generating efficient SQL queries that prune intermediate results and avoid recursion, and (b) using a caching mechanism.}
    \end{figure}
}
\newcommand{\relationalSchema}{
    \begin{table}[t!]
        \centering
        \small
        \caption{Relational schema representation of data model.}
        \vspace{-1em}
        \label{table:relational_schema}
        {\renewcommand{\arraystretch}{1}%
        \begin{tabular}{|l|}
          \hline
          \texttt{Objects}(vid, fid, oid, cid, $x_1$, $y_1$, $x_2$, $y_2$) \\ \hline
          \texttt{Relationships}(vid, fid, rid, $\textrm{oid}_1$, pid, $\textrm{oid}_2$) \\ \hline
          \texttt{Attributes}(vid, fid, oid, key, value) \\ \hline
        \end{tabular}}
    \end{table}
}

\newcommand{\videoSystemComparison}{
    \begin{table}[t!]
        \centering
        \footnotesize
        \caption{Comparison between compositional video analytics systems.}
        \vspace{-1em}
        \label{table:video_system_comparison}
        \setlength{\tabcolsep}{0.15em} %
        {\renewcommand{\arraystretch}{0.5}%
        \begin{tabular}{cccccccccc}
            \toprule
            & \makecell{SVQ++ \\ \cite{DBLP:conf/sigmod/ChaoKX20}} & \makecell{Chen \\ et al.\\~\cite{DBLP:conf/sigmod/Chen0KY21}} & \makecell{Caesar\\~\cite{DBLP:conf/sensys/LiuGUMCG19}} & \makecell{STAR\\~\cite{DBLP:journals/pvldb/ChenKYY22}} & \makecell{VidCEP\\~\cite{DBLP:conf/bigdataconf/YadavC19}} & \makecell{CVQL\\~\cite{DBLP:journals/tmm/KuoC00}} & \makecell{Quivr\\~\cite{mell2021synthesizing}} & \makecell{Rekall\\~\cite{DBLP:journals/corr/abs-1910-02993}} & \textbf{Ours} \\
            \midrule
            \makecell{Object detection} & \checkmark & \checkmark & \checkmark & \checkmark & \checkmark & \checkmark & \checkmark & \checkmark & \checkmark \\
            \makecell{Object tracking} & & \checkmark & \checkmark & \checkmark &  \checkmark & & \checkmark &  \checkmark & \checkmark \\
            \makecell{Relationship} & \checkmark & & \checkmark & \checkmark & \checkmark &\checkmark & \checkmark & \checkmark & \checkmark \\
            \makecell{Attribute} & & & & \checkmark & \checkmark & &  & \checkmark & \checkmark \\  \hline
            \makecell{Conjunction} & & \checkmark & \checkmark & \checkmark & \checkmark & \checkmark & \checkmark & \checkmark & \checkmark \\
            \makecell{Sequencing} & & & \checkmark & \checkmark & \checkmark &\checkmark & \checkmark & \checkmark & \checkmark \\
            \makecell{Iteration} & & \checkmark & &  & & \checkmark & \checkmark &   \checkmark & \checkmark \\
            \makecell{Window} & & \checkmark &  &  & \checkmark & & & \checkmark & \checkmark \\ \hline
            \makecell{Query by example} & & & & & & & \checkmark & & \checkmark \\
            \bottomrule
        \end{tabular}}
        \vspace{-0.6cm}
    \end{table}
}

\newcommand{\trajectoryQueries}{
    \begin{table*}[t!]
        \centering
        \footnotesize
        \caption{Example queries written for the trajectories dataset. ``Pos \%'' is the percentage of positive examples in the dataset.}
        \vspace{-1em}
        \label{table:trajectory_queries}
        {\renewcommand{\arraystretch}{0.8}%
        \begin{tabular}{l p{0.32\textwidth} p{0.52\textwidth} l}
            \toprule
            ID & Query & Description & Pos \% \\
            \midrule
            TQ1 & $(\texttt{Near}(o_1, o_2), \texttt{Bottom}(o_1))$ & $o_1$ is close to $o_2$ while $o_1$ is at the bottom. & 7.0\% \\
            TQ2 & $\texttt{Far}(o_1, o_2); \texttt{Near}(o_1, o_2); \texttt{Far}(o_1, o_2)$ & Two objects move from far to close, then to far again. & 7.5\% \\
            TQ3 & $\texttt{Far}(o_1, o_2)$; $(\texttt{Near}(o_1, o_2), \texttt{Behind}(o_1, o_2))$ & $o_1$ and $o_2$ are far away, then they move close and $o_1$ is behind $o_2$. & 12\% \\
            TQ4 & $\texttt{Far}(o_1, o_2)$; $(\texttt{Near}(o_1, o_2), \texttt{Behind}(o_1, o_2), \texttt{Left}(o_1))$ & $o_1$ and $o_2$ are far apart, then they move close and $o_1$ is behind $o_2$ and $o_1$ is on the left. & 6.5\%\\
            TQ5 & $(\texttt{FrontOf}(o_1, o_2), \texttt{Top}(o_1))$ & $o_1$ is in front of $o_2$ while $o_1$ is at the top. & 45\% \\
            TQ6 & $\texttt{Near}(o_1, o_2); \texttt{Far}(o_1, o_2)$ & Two objects move from close to far apart. & 10\% \\
            TQ7 & $(\texttt{Near}(o_1, o_2), \texttt{Left}(o_1), \texttt{Behind}(o_1, o_2))$ &$o_1$ is close to and behind $o_2$ while $o_1$ is on the left. & 8.5\% \\
            TQ8 & $(\texttt{Far}(o_1, o_2), \texttt{Bottom}(o_1)); \texttt{Near}(o_1, o_2)$ & $o_1$ at the bottom is far from $o_2$, then they move close. & 6.9\% \\
            TQ9 & $(\texttt{Far}(o_1, o_2), \texttt{Left}(o_1)); (\texttt{Near}(o_1, o_2), \texttt{Left}(o_1))$ & $o_1$ and $o_2$ move from far to close while $o_1$ is on the left. & 10\% \\
            TQ1D & $\texttt{Duration}(\texttt{Far}(o_1, o_2), 5); \texttt{Near}(o_1, o_2); \texttt{Far}(o_1, o_2)$ & Two objects are far apart for at least 5 frames, then they move close, then they are far again. & 5.6\% \\
            TQ2D & $\texttt{Duration}(\texttt{LeftOf}(o_1, o_2), 5); (\texttt{Near}(o_1, o_2), \texttt{Top}(o_1))$; $\texttt{Duration}(\texttt{RightOf}(o_1, o_2), 5)$ & $o_1$ is on the left of $o_2$ for at least 5 frames, then they move close to each other and $o_1$ is at the top, then $o_1$ is on the right of $o_2$ for at least 5 frames. & 6\% \\
            TQ3D & $\texttt{Duration}((\texttt{Frontof}(o_1, o_2), \texttt{Left}(o_1)), 15)$; \newline $\texttt{Duration}((\texttt{Left}(o_1), \texttt{RightOf}(o_1, o_2), \texttt{Top}(o_1)), 5)$ & $o_1$ is in front of $o_2$ while $o_1$ is on the left for at least 15 frames, then $o_1$ moves to the right of $o_2$ while $o_1$ is at the top left for at least 5 frames. & 8.3\% \\
            \bottomrule
        \end{tabular}}
        \vspace{-0.6em}
    \end{table*}
}

\newcommand{\warsawQueries}{
    \begin{table*}[t!]
        \centering
        \footnotesize
        \caption{Example queries written for the real-world dataset. ``$\#$ Pos '' is the number of positive examples in the dataset.}
        \vspace{-1em}
        \label{table:warsaw_queries}
        {\renewcommand{\arraystretch}{0.8}%
        \begin{tabular}{l p{0.33\textwidth} p{0.45\textwidth} l}
            \toprule
            ID & Query & Description & $\#$ Pos \\
            \midrule
            WQ1,2,3 & $(\texttt{LaneA}(o_1), \texttt{LaneA}(o_2), \texttt{Near}(o_1, o_2))$ & $o_1$ and $o_2$ are close and in the same lane. & 164, 140, 228\\
            WQ4,5,6 & $(\texttt{LaneA}(o_1), \texttt{HighAccel}(o_1)); (\texttt{LaneA}(o_2), \texttt{HighAccel}(o_2))$ & In lane 1, $o_1$ accelerates rapidly and then $o_2$ accelerates rapdily. & 790, 495, 657\\
            WQ7 & $(\texttt{LaneA}(o_1), \texttt{LaneB}(o_2)); (\texttt{LaneB}(o_1), \texttt{LaneB}(o_2))$ & $o_1$ is turning from lane A into lane B while $o_2$ is in lane B.   & 105 \\
            WQ8,9 & $(\texttt{LaneA}(o_1), \texttt{LaneC}(o_2)); (\texttt{LaneB}(o_1), \texttt{LaneC}(o_2))$ & $o_1$ merges from lane A into lane B while $o_2$ is in lane C (which is next to lane B). & 492, 216\\
            WQ1,2D & $\texttt{Duration}((\texttt{LaneA}(o_1), \texttt{LaneB}(o_2), \texttt{Near}(o_1, o_2)), 5)$ & $o_1$ and $o_2$ are in adjacent lanes and close for at least 5 frames. & 177, 435 \\
            WQ3,4D & $\texttt{Duration}((\texttt{LaneA}(o_1), \texttt{LaneB}(o_2), \texttt{Faster}(o_1, o_2)), 5)$ & $o_1$ and $o_2$ are in adjacent lanes and $o_1$ is faster than $o_2$ for at least 5 frames. & 471, 148\\
            \bottomrule
        \end{tabular}}
        \vspace{-0.6em}
    \end{table*}
}

\newcommand{\userStudyQueries}{
    \begin{table*}[t!]
        \centering
        \footnotesize
        \caption{Queries used in the user study. ``Pos \%'' is the percentage of positive examples in the dataset.}
        \vspace{-1em}
        \label{table:user_study_queries}
        \setlength{\tabcolsep}{0.3em} %
        {\renewcommand{\arraystretch}{0.8}%
        \begin{tabular}{l p{0.38\textwidth} p{0.52\textwidth} l}
            \toprule
            ID & Query & Description & Pos $\%$ \\
            \midrule
            UQ1 & $(\texttt{Color}(o_1, \textrm{`red'}), \texttt{Shape}(o_2, \textrm{`cylinder'}), \texttt{Far}(o_1, o_2)); \texttt{Near}(o_1, o_2)$ & A red object is far from a cylinder, then they get close. & $16.7\%$\\
            UQ2 & $(\texttt{Color}(o_1, \textrm{`purple'}), \texttt{Material}(o_1, \textrm{`metal'}),$ \newline $\texttt{Behind}(o_1, o_2), \texttt{Bottom}(o_2))$ & A purple metal object $o_1$ is behind another object $o_2$ at the bottom of the screen. & $24.2\%$\\
            UQ3 & $(\texttt{Color}(o_1, \textrm{`red'}), \texttt{Shape}(o_2, \textrm{`cylinder'}), \texttt{Far}(o_1, o_2));$ \newline $(\texttt{Near}(o_1, o_2), \texttt{Top}(o_3), \texttt{Right}(o_3))$ & A red object is far from a cylinder, then they get close while a third object is at the top right of the screen. & $12.8\%$ \\
            UQ4 & $(\texttt{Color}(o_1, \textrm{`purple'}), \texttt{Material}(o_1, \textrm{`metal'}),$ \newline $\texttt{Behind}(o_1, o_2), \texttt{Bottom}(o_2)); \texttt{Top}(o_2)$ & A purple metal object $o_1$ is behind another object $o_2$ at the bottom of the screen, then $o_2$ moves to the top. & $12.8\%$ \\
            UQ5 & $\texttt{Duration}((\texttt{Color}(o_1, \textrm{`red'}), \texttt{Shape}(o_2, \textrm{`cylinder'}),$ \newline $\texttt{Far}(o_1, o_2)), 25); (\texttt{Near}(o_1, o_2), \texttt{Top}(o_3), \texttt{Right}(o_3))$ & A red object is far from a cylinder for at least a second, then they are near each other while a third object is at the top right of the screen. & $4.2\%$\\
            UQ6 & $(\texttt{Color}(o_1, \textrm{`purple'}), \texttt{Material}(o_1, \textrm{`metal'}), \texttt{Behind}(o_1, o_2),$ \newline $\texttt{Bottom}(o_2)); \texttt{Top}(o_2); \texttt{Duration}((\texttt{Bottom}(o_3), \texttt{Right}(o_3)), 25)$ & A purple metal object $o_1$ is behind another object $o_2$ at the bottom of the screen, then $o_2$ moves to the top, then a third object $o_3$ is at the bottom right of the screen for at least a second. & $3.8\%$\\
            \bottomrule
        \end{tabular}}
        \vspace{-1.2em}
    \end{table*}
}

\newcommand{\sceneGraphQueriesTemplate}{
    \begin{table}[t!]
        \centering
        \small
        \caption{Query configurations for the scene graphs dataset.}
        \label{table:scene_graph_queries_template}
        \setlength{\tabcolsep}{0.5em} %
        {\renewcommand{\arraystretch}{0.7}%
        \begin{tabular}{ccccc}
            \toprule
            Query & relationship \& state  & property  & region graph & duration \\ \midrule
            Easy & 3 & 1 & 1 & No\\
            Medium & 5 & 2 & 3 & No\\
            Hard & 5 & 2 & 3 & Yes\\ \bottomrule
            \end{tabular}
            }
    \end{table}
}
\newcommand{\mainResult}{
    \begin{table*}[t!]
        \centering
        \small
        \caption{Query synthesis time (median, in seconds) for each method to achieve at least 0.9 F1 scores. \system successfully learns all queries with at least 0.9 F1 scores and is faster than (or at least comparable to) the two baselines. (NA: not applicable, ---: failed due to insufficient F1 score or timeout)}
        \label{table:main_result}
        \setlength{\tabcolsep}{0.4em} %
        {\renewcommand{\arraystretch}{1}%
        \begin{tabular}{|l|lllllllll|llllllllllll|}
            \hline
            \multirow{2}{*}{Method} & \multicolumn{9}{c|}{Simplified} & \multicolumn{12}{c|}{Normal} \\
            & TQ1 & TQ2 & TQ3 & TQ4 & TQ5 & TQ6 & TQ7 & TQ8 & TQ9 & TQ1 & TQ2 & TQ3 & TQ4 & TQ5 & TQ6 & TQ7 & TQ8 & TQ9 & TQ1D & TQ2D & TQ3D \\ \hline
            PATSQL & \textbf{1.54} & --- & 538 & --- & \textbf{4.00} & 1.93 & 552 & --- & --- & NA & NA & NA & NA & NA & NA & NA & NA & NA & NA & NA & NA \\
            Quivr & 10.5 & 5.77 & 27.9 & 94.2 & 13.5 & 4.36 & 47.7 & \textbf{16.7} & --- & 7228 & 7428 & 8185 & --- & 7839 &  3683 & --- & 8756 & --- & --- & --- & --- \\
            Ours & 3.57 & \textbf{1.85} & \textbf{17.1} & \textbf{46.6} & 4.36 & \textbf{1.66} & \textbf{18.5} & 22.7 & \textbf{20.6} & \textbf{75.6} & \textbf{125} & \textbf{106} & \textbf{185} & \textbf{50.4} & \textbf{61.1} & \textbf{106} & \textbf{110} & \textbf{107} & \textbf{223} & \textbf{166} & \textbf{183} \\ \hline
        \end{tabular}}
    \end{table*}
}

\newcommand{\quivrNoisyData}{
    \begin{table}[t!]
        \centering
        \small
        \caption{Probability that a system returns at least one query on noisy data. (false positive rate is 0.1 of false negative rate)}
        \label{table:quivr_noisy_data}
        \setlength{\tabcolsep}{0.2em} %
        {\renewcommand{\arraystretch}{1}
        \begin{tabular}{|c|ccccc|}
            \hline
            FN rate & 0.1 & 0.2 & 0.3 & 0.4 & 0.5 \\ \hline
            \makecell{Quivr, Mean \\ (Range)} & \makecell{68\% \\ (20\%-100\%)} & \makecell{54\% \\ (35\%-75\%)} & \makecell{47\% \\ (5\%-75\%)} & \makecell{34\% \\ (5\%-55\%)} & \makecell{17\% \\ (0\%-50\%)} \\ \hline
            \makecell{Ours, Mean} & 100\% & 100\% & 100\% & 100\% & 100\% \\ \hline
        \end{tabular}
        }
    \end{table}
}

\newcommand{\queryExplored}{
    \begin{table}[t!]
        \caption{Number of queries explored and number of predictions per second to achieve at least 0.9 F1 scores. \system reduces computational effort by exploring fewer queries and by executing queries faster.}
        \label{table:query_explored_and_predictions_per_second}
        \parbox{.45\linewidth}{
        \centering
        \small
        \begin{tabular}{|c|c|c|}
            \hline
            \multirow{2}{*}{ID} & \multicolumn{2}{c|}{\# queries explored} \\ \cline{2-3}
             & Quivr & Ours \\ \hline
            TQ1 & 222748 & 1123 \\
            TQ2 & 245112 & 1071 \\
            TQ3 & 220627 & 1101 \\
            TQ5 & 212066 & 1118 \\
            TQ6 & 224504 & 1077 \\
            TQ8 & 263040 & 1080 \\ \hline
        \end{tabular}
        \label{table:query_explored}
        }
        \hfill
        \parbox{.45\linewidth}{
        \centering
        \small
        \begin{tabular}{|c|c|c|}
            \hline
            \multirow{2}{*}{ID} & \multicolumn{2}{c|}{\# predictions/s} \\ \cline{2-3}
            & Quivr & Ours \\ \hline
            TQ1 & 154 & 574 \\
            TQ2 & 124 & 666 \\
            TQ3 & 140 & 664 \\
            TQ5 & 140 & 480 \\
            TQ6 & 151 & 547 \\
            TQ8 & 145 & 792 \\ \hline
        \end{tabular}
        \label{table:predictions_per_second}
        }
        \vspace{1em}
    \end{table}
}

\newcommand{\warsawSuccessRate}{
    \begin{table}[t!]
        \centering
        \small
        \caption{Quivr's success rate on the real-world dataset.}
        \label{table:warsaw_success_rate}
        \setlength{\tabcolsep}{0.5em} %
        {\renewcommand{\arraystretch}{1}%
        \begin{tabular}{|c|cccccc|}
            \hline
            Query & WQ1 & WQ4 & WQ7 & WQ8 & WQ1D & WQ3D \\ \hline
            Quivr success rate & 40\% & 55\% & 35\% & 60\% & 5\% & 5\% \\ \hline
            \end{tabular}
            }
    \end{table}
}

\newcommand{\userStudyErrorRate}{
    \begin{table}[t!]
        \centering
        \small
        \caption{Labeling error rates of participants in the user study.}
        \label{table:user_study_error_rate}
        {\renewcommand{\arraystretch}{1}
        \begin{tabular}{|c|cccccc|}
            \hline
            Error type & UQ1 & UQ2 & UQ3 & UQ4 & UQ5 & UQ6 \\ \hline
            FP rate & 0.250 & 0.065 & 0.169 & 0.028 & 0.308 & 0.029\\
            FN rate & 0.061 & 0.123 & 0.091 & 0.133 & 0.286 & 0.714\\ \hline
        \end{tabular}}
        \vspace{1em}
    \end{table}
}

\newcommand{\machineLearningMethods}{
    \begin{table}[t!]
        \centering
        \small
        \caption{Median F1 scores of \system and ML methods. \system achieves higher F1 scores for all queries.}
        \label{table:machine_learning_methods}
        {\renewcommand{\arraystretch}{1}
        \begin{tabular}{|c|ccccc|}
            \hline
            Queries & Ours & \textsc{CLIP} & \textsc{MViT} & \textsc{CLIP} (all) & \textsc{MViT} (all) \\ \hline
            Easy & \textbf{0.997} & 0.249 & 0.222 & 0.327 & 0.322 \\
            Medium & \textbf{0.843} & 0.201 & 0.182 & 0.278 & 0.255 \\
            Hard & \textbf{0.638} & 0.219 & 0.203 & 0.286 & 0.265 \\ \hline
        \end{tabular}}
        \vspace{1em}
    \end{table}
}

\section{Introduction}

Video data is increasingly becoming a prized commodity. Inexpensive large-scale video storage and advances in machine learning and computer vision have propelled the use of large video datasets with new applications including drone analytics~\cite{DBLP:conf/edge/WangFCGBPYS18,DBLP:conf/mobicom/WangCC16}, citywide traffic analytics~\cite{DBLP:journals/computer/Ananthanarayanan17, DBLP:conf/intellisys/HammerLFLW20}, civil engineering~\cite{DBLP:conf/wmcsa/GeorgeWBEPS19,barmpounakis2016unmanned}, and many others~\cite{10.1145/3197517.3201394,geiger2012we,senior2007video,DBLP:journals/tvcg/SteinJLBZGSAGK18,mady2021bird}.
Although video database management systems (VDBMSs) have recently re-emerged as an active research area to support these applications~\cite{DBLP:journals/pvldb/KangEABZ17,DBLP:conf/icde/AndersonCRW19,DBLP:journals/pvldb/KangBZ19,DBLP:conf/sigmod/BastaniHBGABCKM20,DBLP:conf/sensys/LiuGUMCG19,DBLP:journals/corr/abs-1910-02993,DBLP:journals/corr/abs-2104-06142}, existing systems fall short of supporting many use cases.

Consider a traffic analytics application: A traffic engineer may want to understand road hazards involving car and motorcycle
interactions (e.g., motorcycles swerving abruptly in front of turning cars). Although many computer vision models exist that detect common objects (e.g.,~``cars'' and ``motorcycles'')~\cite{Wojke2018deep} and relate objects spatially (e.g.,~``bottom of'', ``left of'', ``near'')~\cite{Chen2019SceneGP}, a specific classifier that identifies ``a motorcycle swerving in front of a car, while the car is turning at an intersection'' is unlikely to exist~\cite{DBLP:conf/cvpr/YouJWFL16, DBLP:conf/bigdataconf/YadavC19}. Worse, training one would require many hours of user effort in labeling for a single query. Given the relative rarity of most interesting events, finding sufficient positive instances further exacerbates these labeling requirements. In our example, there will
be many instances of cars and motorcycles in intersections. Only rarely would a motorcycle swerve in front of a turning car.

\systemTeaserFigure

Assuming that we can run existing computer vision models on videos to identify objects, extract attributes, and reason about their pairwise relationships, some recent video data management systems support users by providing an interface to express a declarative query as a composition of extracted information~\cite{DBLP:journals/corr/abs-1910-02993, DBLP:conf/sigmod/ChaoKX20, DBLP:conf/sensys/LiuGUMCG19, DBLP:journals/pvldb/ChenKYY22, DBLP:conf/bigdataconf/YadavC19, mell2021synthesizing, DBLP:journals/pvldb/BastaniMM20}. For example, a user might be able to query for an event using a specification that searches for video clips containing \texttt{car} and \texttt{motorcycle} objects, and specifying their desired relationships (\texttt{near} then \texttt{front\_of}, etc.).
These systems expect users to possess a level of database expertise to be able to express such queries. Additionally, real-world events can be difficult to express declaratively---even for experts. For example, there are multiple ways to express our sample query. The best way depends on the data (e.g., in a given video, motorcycles may be swerving from outside the frame, as illustrated in \Cref{fig:exampleFrames}, discussed later). Other similar systems ask users to sketch their events~\cite{DBLP:journals/pvldb/ChenKYY22, DBLP:conf/mm/ChangCMSZ97}; this is equally challenging for the same reason: an event of interest may deviate from the exact user-provided sketch.

In this paper, we present \textbf{\system} (\Cref{fig:system_teaser}), a system that addresses the above challenge by synthesizing declarative queries on behalf of users from a small number of labeled video segments. Put another way, \system \textbf{E}volves \textbf{Q}ueries for \textbf{U}sers \textbf{I}teratively and is part of our larger \textbf{VOCAL} system~\cite{DBLP:conf/cidr/DaumZHBHKCW22}. In our example, a user provides as few as two positive and ten negative examples containing the event of interest (see \Cref{sec:evaluation}). \system then synthesizes a declarative query and executes it on the remaining large pool of video data to identify other examples of the desired event.

\system supports users' search for complex events without requiring any specific domain knowledge, with limited effort, and without requiring precise declarative specifications or sketches. Similar to other systems, we use the insight that while most user queries are new and unseen, they are usually composed of known, previously seen atoms. Computer vision models for common atoms already exist such as objects (e.g.,~``car'', ``backpack'') and their spatial and semantic relationships (e.g.,~``left of'', ``holding''). Formally, such a composition of visual scenes is referred to as a scene graph in the computer vision community~\cite{DBLP:journals/ijcv/KrishnaZGJHKCKL17,ji2020action} and was developed from its cognitive grounding in human perception~\cite{zacks2001perceiving,kurby2008segmentation,biederman1987recognition}. Key to \system's contribution
is to encapsulate spatio-temporal scene graphs in its data model and use them to define a query language. A spatio-temporal scene graph conceptualizes the contents of a video as a sequence of graphs, each graph representing a single video frame. Each graph contains vertices, which represent objects in the frame; edges represent the relationships between those objects. Each object can possess a set of attributes that describe its properties (e.g., ``red'', ``leather'').
\system extracts relevant data from each video using user-provided functions: i.e.,~pre-existing detectors and classifiers; it synthesizes queries as a composition of extracted scene graph atoms.
We show that \system's data model and query language, both based on the relational model, can express a variety of compositional queries.

Leveraging scene graphs, \system contributes a new query synthesis approach that finds user events with far fewer labeled examples than would be required to train a specialized machine learning model directly, and that works on noisy, video-scale data, and complex events. To support query synthesis in such environment, \system solves two technical challenges: it reduces \textbf{computational effort} and \textbf{user effort}.

\system reduces \textbf{computational effort} by limiting query search using scene graphs, by pruning search paths using beam search, and by avoiding expensive database operations.
First, unlike prior query-by-example techniques that synthesize arbitrary SQL~\cite{DBLP:journals/pvldb/FarihaM19, DBLP:conf/sigmod/PsallidasDCC15, DBLP:journals/pvldb/LiCM15, Wang2017SynthesizingHE, DBLP:journals/pvldb/TakenouchiIOS21}, \system reduces the search space by limiting the query search to sequences of scene graphs.
Second, synthesizing queries over sequences of scene graphs can still be a computationally slow process to traverse the search space of possible queries. Existing query-by-example systems enumerate all possible queries; although pruning techniques like equivalence classes~\cite{DBLP:conf/pldi/UdupaRDMMA13}, over-approximation~\cite{Wang2017SynthesizingHE}, and lifting projection operators~\cite{DBLP:journals/pvldb/TakenouchiIOS21} can be used to reduce the search space, these mechanisms are not sufficient to make exhaustive exploration tractable. Instead, \system adopts a beam search strategy to explore the query space efficiently. Beam search limits exploration to a subset of the most promising branches at each step.
Third, executing the many candidate queries on the user examples is prohibitively expensive. Existing systems~\cite{Wang2017SynthesizingHE, DBLP:journals/pvldb/TakenouchiIOS21} evaluate candidate queries with many joins and thus do not scale well when the size of user examples becomes large. \system carefully generates efficient queries that avoid expensive operations such as recursive joins. \system comes with a set of optimizations to generate efficient SQL statements and uses the PostgreSQL database engine to execute them.

\system reduces \textbf{user effort} by using active learning and by being robust to noise.
With active learning, \system reduces the number of labeled examples needed: Instead of asking a user to provide all examples up front, \system iteratively requests labels of carefully selected additional examples to reduce the uncertainty in query synthesis.
Noise can naturally creep into systems that interface with user labeling, machine learning models, and potentially ambiguous real-world events. Distinct from other existing systems~\cite{DBLP:journals/pvldb/TakenouchiIOS21,mell2021synthesizing}, \system searches for queries that best match potentially noisy data and input. It also retains imperfect query candidates at every iteration and uses regularization to prevent overfitting to noise or limited user input.

In summary, \system makes the following contributions:
 \begin{itemize}
     \item We introduce an expressive data model and a query language based on spatio-temporal scene graphs (\Cref{sec:datamodel}).
     \item We propose a new approach that efficiently synthesizes the user's intended query from examples. We limit the query search space using scene graphs, prune search paths using beam search, leverage active learning to reduce user effort, and handle noisy data (\Cref{sec:synthesis}).
     \item We implement a set of optimizations to generate efficient SQL query statements and reduce computational effort during query synthesis (\Cref{sec:query-execution}).
     \item We evaluate our approach on synthetic and real datasets~\cite{CLEVRER2020ICLR, DBLP:conf/sigmod/BastaniHBGABCKM20} and show that it outperforms two baselines~\cite{DBLP:journals/pvldb/TakenouchiIOS21,mell2021synthesizing}---in terms of F1 score, synthesis time, and robustness to data noise---and can flexibly synthesize complex queries that the baselines do not support. We also conduct a user study to show the performance of real users using \system. (\Cref{sec:evaluation}).
 \end{itemize}

 Overall, \system is an important step toward making video database management systems more accessible to experts and non-experts alike, by easing the task of expressing queries over videos.

\section{\system data model}\label{sec:datamodel}

\exampleFramesFigure

This section describes \system's data model and query language, which
we briefly introduced in our vision paper~\cite{DBLP:conf/cidr/DaumZHBHKCW22}, but develop in depth here. \Cref{sec:synthesis} shows how \system synthesizes queries from user input using this data model.

For ease of presentation, we use a simplified, running example, where a traffic engineer seeks to find instances of ``a car arriving from the left and passing a motorcycle at the intersection.'' \Cref{fig:exampleFrames} shows two representative frames from a video that contains such an event.
We show other example queries in \Cref{sec:evaluation}.

\subsection{Scene graphs as our data model}\label{subsec:datamodel}

\system represents a video $V$ as a set of short, non-overlapping \textit{video segments}, $v \in V$ (5-second segments in our prototype implementation).
Each video segment is a sequence of $N$ frames $\{f_1,\ldots, f_N\}$. The visual content of each frame is represented by a scene graph~\cite{DBLP:journals/ijcv/KrishnaZGJHKCKL17}:
A \textit{scene graph} $g_i= (\textbf{o}_i, \textbf{r}_i)$ contains the set of all objects $\textbf{o}_i$ in a frame, along with a set of all relationships $\textbf{r}_i$ between those objects. Often a scene graph contains more information than is necessary to identify an event, and so the literature also defines a \textit{region graph} $g_{ij}$, which is a subgraph of $g_i$, i.e., $g_{ij} \subseteq g_i$. \Cref{fig:exampleFrames} (left) shows an example frame and region graph.

We define an \textit{object} in a frame as $o = (vid, f_i, oid, cid, bbox)$, where $f_i$ is the sequence number of frame $i$ in video segment, $vid$. $oid$ is a unique identifier of the object in the video segment, $cid$ is the identifier for the class of the object (e.g., ``car'', ``motorcycle''), and $bbox$ is the bounding box containing the object in frame, $f_i$.  A $bbox$ is represented by its upper-left and bottom-right coordinates, i.e., $bbox = (x_1, y_1, x_2, y_2)$.

Objects can have intra-frame \textit{relationships} defined as $ r = (vid, f_i, rid, oid_{\textrm{sub}}, pid, oid_{\textrm{tar}})$, where $rid$ is a unique identifier of the relationship in frame, $f_i$. Subject, $o_{\textrm{sub}}$, is connected to target, $o_{\textrm{tar}}$, with the relationship class identifier, $pid$ (e.g., ``near'' or ``holds'').
Both subject and target belong to frame $f_i$: i.e., $o_{\textrm{sub}}, o_{\textrm{tar}} \in \textbf{o}_i$.

Objects can have \textit{attributes} $a = (vid, f_i, oid, k, v)$, where $k$ is the name of the attribute, $v$ is the value of the attribute, $vid$, $f_i$ and $oid$ identify the video segment, frame, and object. \system distinguishes \textit{state} and \textit{property} attributes. The former change over time and are typically computed from the bounding box of an object (e.g., ``location=bottom''). The latter capture intrinsic properties of objects and are immutable (e.g., ``color=red'').

Finally, an \textit{event} $e$ is a temporally ordered sequence of region graphs $ e = (eid, \{g_1, \ldots, g_{k}\}) $. Region graphs in an event do not need to be contiguous or distinct.

\subsubsection*{Example} Suppose that the two frames in~\Cref{fig:exampleFrames} are the 10th and 15th frames of a video segment \texttt{V1}, and that the motorcycle and car are the 7th and 9th objects detected in \texttt{V1}. Then, for the left frame, we generate the region graph $g_1=(\textbf{o}_1, \textbf{r}_1)$, where:
  $\textbf{o}_1=\{o_{11}, o_{12}\}$ represents the car $o_{11} = (\texttt{V1}, \texttt{F10}, \texttt{OID9}, \textrm{car}, bbox_1)$ and the motorcycle $o_{12}=(\texttt{V1}, \texttt{F10}, \texttt{OID7}, \textrm{motorcycle}, bbox_2)$ and
  $\textbf{r}_1=\{r_{11}\}$ contains a relationship $r_{11}=(\texttt{V1}, \texttt{F10}, \texttt{RID1}, \texttt{OID9}, \textrm{leftOf}, \texttt{OID7})$. The
  car also has an attribute $a_{11}=(\texttt{V1}, \texttt{F10}, \texttt{OID9}, \textrm{location}, \textrm{bottom})$. We can define the region graph $g_2$ for the right frame similarly. The only difference will be the relationship between the object will indicate that the car is now \textrm{rightOf} the motorcycle. Finally, the event $e = (\texttt{EID1}, \{g_1, g_2\})$ represents a car arriving from the left and passing a motorcycle at the intersection.

The relational schema in~\Cref{table:relational_schema} captures the above data model.
The benefit of using a relational schema is that we can execute relational queries to specify region graphs and find events of interest, which is flexible and follows the well-understood semantics of the relational model.
For each video (or collection of related videos), \system creates a \textit{view} with this schema. In~\Cref{sec:query-execution} we describe when and how relations in the view are materialized.

\relationalSchema

To populate each relation, \system uses available user-defined functions (\autoref{fig:system_teaser}). User-defined functions can be provided by the user or be available publicly in the form of existing machine learning models, such as object detectors. Various user-defined functions can be declared in \system: (i) an object detector~\cite{DBLP:conf/nips/RenHGS15} that takes a video frame as input and outputs the set of objects and their bounding boxes, (ii) an object tracking algorithm~\cite{Wojke2018deep} that takes objects in consecutive frames as input and, for each pair of objects, determines if they are the same, and (iii) a set of pre-trained models (e.g.,~\cite{Chen2019SceneGP}) or rules that can take two objects in the same frame as input and determine their relationship (e.g., ``near'', ``behind'', ``riding'', ``holding'') or that can take one object as input and determine its attributes (e.g., ``location=bottom'', ``color=red''). In our experiments, \system uses a general-purpose object detector~\cite{DBLP:conf/iccv/HeGDG17} to locate objects and intrinsic attributes of objects to generate trajectories across frames~\cite{CLEVRER2020ICLR}.

\subsection{Scene graphs as our query language}\label{subsec:query_language}

\system could execute arbitrary relational queries on the view defined above. However, this would form an intractable search space, making query synthesis unusably slow for most real-world applications. Instead, we define a query language that is more restrictive, affording a smaller search space and therefore, faster synthesis. We constrain queries to (i) a temporally ordered sequence of region graphs, (ii) a set of predicates, (iii) a set of duration constraints, (iv) a window specification, and (v) to output video
segment identifiers. Using Datalog and with some abuse of notation, a query in \system can be expressed as:
$$ q(vid) \textrm{ :- } g_1, \ldots, g_k, \textbf{p}, \textbf{d}, w\text{, where:}$$
\begin{itemize}[leftmargin=2em]
    \item A temporally ordered sequence of region graphs $g_1, \ldots, g_k$ specifies that a matching event consists of $g_1$, followed by $g_2$, followed by $g_3$, etc. Each $g_i$ is specified with a set of atoms: \texttt{Objects}, \texttt{Relationships}, and \texttt{Attributes} joined on a shared $vid$ and $fid$. Moreover, each $g_i$ can persist for multiple frames and there can be other frames between $g_i$ and $g_{i+1}$.
    \item A set of predicates $\textbf{p}$ can be applied to objects, relationships, and attributes. In our example, predicates would specify that the query is looking for ``car'' and ``motorcycle'' objects, that the car needs to be ``left of'' then ``right of'' the motorcycle, and that the car should be at the ``bottom'' of the frame.
    \item  A set of duration constraints $\textbf{d}$ can be applied to region graphs and define the minimum number of contiguous frames that a region graph $g_i$ should be valid before transitioning to the next region graph $g_{i+1}$.
    \item A window specification $w$ is the maximum number of frames that can separate $g_1$ from $g_{k}$.
    \end{itemize}

    For example, following the above restricted template, the event in \Cref{subsec:datamodel} can be expressed with the following Datalog rules:\footnote{In this and the following examples we use English words instead of integers for $cid$ and $pid$ values to make the examples more readable}

    \begingroup
    \fontsize{7pt}{9pt}\selectfont
    \begin{verbatim}
  g1(vid, fid, oid1, oid2) :-
      Objects(vid, fid, oid1, 'car', _, _, _, _),
      Objects(vid, fid, oid2, 'motorcycle', _, _, _, _),
      Relationships(vid, fid, _, oid1, 'leftOf', oid2),
      Attributes(vid, fid, oid1, 'location', 'bottom'), oid1 != oid2.
    \end{verbatim}%
    \endgroup

The above rule will find frames in video segments that contain a car and a motorcycle, such that the car is on the left of the motorcycle, and the car is at the bottom of the frame, where the intersection is located. Next, the event will likely
consist of a sequence of such frames, which can be captured with the following recursive rules:

\begingroup
    \fontsize{7pt}{9pt}\selectfont
    \begin{verbatim}
  g1_star(vid, fid, fid, oid1, oid2) :- g1(vid, fid, oid1, oid2).
  g1_star(vid, fid_start, fid_end, oid1, oid2) :-
      g1_star(vid, fid_start, fid, oid1, oid2),
      g1(vid, fid_end, oid1, oid2), fid_end = fid + 1.
    \end{verbatim}%
    \vspace{-1.5em}
\endgroup%

We could use equivalent rules to define \texttt{g2} and \texttt{g2\_star}.
Finally, the query that returns matching video segments takes the form:

\begingroup
    \fontsize{7pt}{9pt}\selectfont
    \begin{verbatim}
  q(vid) :- g1_star(vid, fid11, fid12, oid1, oid2),
      g2_star(vid, fid21, fid22, oid1, oid2),
      fid21 > fid12, fid22 - fid11 < 1800
  \end{verbatim}%
  \vspace{-1.5em}
\endgroup%

    The predicate $fid21 > fid12$ indicates that the second sequence of region graphs should follow the first one. The predicate
    allows for a gap between sequences, which may arise, for example, if something obstructs the vehicles from the camera's view.
    Finally, $fid22 - fid11 < 1800$ puts a time constraint on the event (a 30-second time-window, assuming 60 frames per second).

To summarize, \system's query language is a subset of Datalog with recursion, expressed on a specific schema.
In~\Cref{sec:query-execution}, we explain how we avoid executing expensive recursive queries.

\subsection{Expressiveness of our data model}\label{subsec:expressiveness}

\videoSystemComparison

We compare the expressiveness of \system's data model against other compositional video analytics systems in~\Cref{table:video_system_comparison}. Among them, SVQ++~\cite{DBLP:conf/sigmod/ChaoKX20} only supports spatial relationships between two objects, Chen et al.~\cite{DBLP:conf/sigmod/Chen0KY21} only supports temporal queries that count co-occurring objects. Caesar~\cite{DBLP:conf/sensys/LiuGUMCG19}, STAR Retrieval~\cite{DBLP:journals/pvldb/ChenKYY22}, and VidCEP~\cite{DBLP:conf/bigdataconf/YadavC19} lack the important feature of \textit{iteration} that allows a region graph to persist for multiple frames and thus cannot support duration constraints. CVQL~\cite{DBLP:journals/tmm/KuoC00} does not track objects and all its predicates are defined at the object class level.
Quivr~\cite{mell2021synthesizing} has similar expressiveness to us, but is limited to trajectory queries. Rekall~\cite{DBLP:journals/corr/abs-1910-02993} introduces a flexible Python library for video event specification, but requires that users manually write and refine queries.

Our data model currently does not support tertiary relationships between three objects, such as ``a person hitting a ball with a bat''. The closest approximation we have is ``a person holding a bat'' and ``a bat hitting a ball''. We also do not support disjunctions or negations.
Several existing systems have support for these operators. For example, Chen et al.~\cite{DBLP:conf/sigmod/Chen0KY21} supports arbitrary CNF queries, Caesar~\cite{DBLP:conf/sensys/LiuGUMCG19} supports disjunctions, and Rekall~\cite{DBLP:journals/corr/abs-1910-02993} can flexibly support all three operators. However, the focus of these systems lies in query execution, rather than in learning query specifications from examples, which is the primary focus of this paper.

Datalog and scene graphs serve as the foundation of our query language. By building on the relational model and Datalog, we ensure that our queries have precise and well-understood semantics. By focusing on queries that specify a desired sequence of region graphs, we constrain the search space for queries, which accelerates synthesis, we are able to optimize the execution of those queries (compared with trying to support arbitrary recursive Datalog queries), and we are still able to capture events that occur in videos, where objects interact with each other in space and time.

\section{Query synthesis}~\label{sec:synthesis}

With our data model and query language defined, we now formally present the query synthesis problem statement, then describe all the components of our proposed solution.

\vspace{-0.2em}
\subsection{Query synthesis problem statement}
Following the aforementioned data model, a user would like to execute a query $q_u$ on video database $D$ that returns a set of video segment identifiers, $V_{output}=q_u(D)$.
Given the user's intended query, $q_u$, each video segment $v_i$ can be seen as having a ground-truth label $y_i \in \{0,1\}$ indicating whether it matches $q_u$.
Initially, both the ground-truth labels and $q_u$ are unknown because the user is unable to specify $q_u$ and can only label video segments as positive or negative instances.
The goal of \system is to synthesize a target query, $q_t \in Q$, that is the best approximation of $q_u$ in its search space. When executed over database $D$, $q_t$ should yield the best measure performance (e.g., F1 score):
$$ q_t = \arg\max_{q\in Q} \textrm{measure}(q(D), q_u(D)) $$
\system can request a label from the user $O$: $\hat{y_i}=O\left(v_i\right)$. Since user labels may be noisy, it is possible that $\hat{y_i} \neq y_i$. Given that enumerative search is intractable, \system uses a heuristic approach to find an approximately-best query $\hat{q}_t$ to the objective, while reducing both user effort and query synthesis time.

\subsection{Query synthesis algorithm overview}~\label{subsec:algorithm}

\setlength{\textfloatsep}{0pt}%
\begin{algorithm}[t!]
    \small
    \setstretch{0.8}
    \SetNoFillComment
    \SetKwInOut{Input}{Input}
    \SetKwInOut{Output}{Output}

    \Input{$U$ - set of unlabeled video segments \\
      $L$ - set of labeled video segments \\
   $P$ - set of user-defined functions \\
    $b$, $bw$, $s_q$, $s_v$, $k$ - hyperparameters \\
    }
    \Output{$Q_t$ - set of top-$k$ target queries}
    $S \gets \{q_\emptyset\}$ \\
    \While{$|S| > 0$}
    {
        $S' \gets \{\}$ \label{line:expand_start} \\
        \ForEach{$q \in S$} {
            $S' \gets S' \cup \textsc{ExpandQuery}(q, P)$ \label{line:expand_end}
        }
        \If{$|L| < b$} {
            $L, U \gets \textsc{PickNextSegments}(L, U, S')$ \label{line:pick_next}
        }
        $S \gets \textsc{SampleQueries}(S', L, bw)$ \\ \label{line:prune}
        $Q_t \gets Q_t \cup S'$ \\
        $Q_t \gets \textsc{RetainTopQueries}(Q_t, L, k)$ \\ \label{line:retain_top}
    }
    \Return{$Q_t$}
    \caption{Query synthesis algorithm that returns top-$k$ synthesized queries matching user input.}\label{algorithm}
\end{algorithm}

Algorithm~\ref{algorithm} provides an overview of \system's synthesis algorithm. It traverses the space
of possible queries to synthesize a final target query that matches the user's intent.
The algorithm takes as input the set of unlabeled video segments, $U$, and a small set of labeled segments, $L$.
$L$ is provided by the user and should include both positive and negative examples of the desired event.
In our experiments, we require as few as two positive and ten negative examples. While providing two positive examples requires some work from the user, such a setting is common in query-by-example approaches~\cite{mell2021synthesizing, DBLP:journals/pvldb/FarihaM19, DBLP:conf/popl/RaychevBVK16}. We leave starting with zero examples for future work.
The algorithm also takes as input a set of user-defined functions, $P$, comprised
of indicator functions for object classes (e.g., “car”), relationships (e.g., “near”), and attribute key-value pairs (e.g., “color=red”). Algorithm~\ref{algorithm}
enumerates $P$ when expanding its search space.

The algorithm returns a set of target queries, $Q_t$, which is the top-$k$ synthesized queries ranked by a performance measure (F1 score in our prototype implementation) and is updated regularly through the search process. The learned queries can be applied to unseen videos to find video segments containing the matching event. By default, \system returns the top-$k$ queries but then executes the best one over unseen videos. The user can also randomly sample one query from $Q_t$, execute multiple queries from $Q_t$ and aggregate the results, or manually examine $Q_t$ to pick the intended query.
If the user is not satisfied with the synthesized query, they can edit it directly. The user can also restart the search with a larger number of initial examples by reusing the video segments labeled during the last session. Moreover, the user could provide a starter query to \system (e.g., a simple query or an earlier query) rather than starting from an empty query.

To populate $Q_t$ efficiently, \system synthesizes queries in a bottom-up fashion. It starts with an empty query, $q_\emptyset$, and incrementally adds predicates to it, up to a certain complexity ($\textsc{ExpandQuery}$ method on line 5 and \Cref{sec:expansion}).
At each step, different predicates can be added to a query, expanding the search in multiple directions. To limit the size of the search space and ensure fast query synthesis, \system adopts a beam search-style strategy ($\textsc{SampleQueries}$ method on line 8 and \Cref{sec:pruning}).

An important challenge for \system is that learning a query from a small number of user-provided examples is difficult. To address this challenge, \system uses active learning to effectively guide the query synthesis process and identify good target queries with limited initial and additional user effort. Specifically, method $\textsc{PickNextSegments}$ on line 7 and in \Cref{sec:active_learning} uses active learning to select additional video segments for the user to label in order to effectively differentiate between multiple candidate subqueries, and expand the search in the most promising directions.

At the end of each iteration, method $\textsc{RetainTopQueries}$ (line 10 and \Cref{sec:retaining_topk_queries}) maintains a list of top-$k$ queries seen so far, which is larger than the number of queries selected for additional expansion, in case a query seen earlier in the search
ends up with the best score on the final labeled set of segments, or, as mentioned above, to give users options if they would like to try
alternative, high-performing queries.

The algorithm has several hyperparameters, including a labeling budget $b$ (i.e., the maximum number of labels the user is willing to provide), the number of candidate queries to retain during exploration $bw$ (i.e., the beam width), the number of candidate queries, $s_q$, and the number of candidate video segments, $s_v$, sampled during active learning, and the number of queries in the final answer $k$. \Cref{fig:algorithmExample} illustrates the algorithm using the running example.

\algorithmExampleFigure

\vspace{-1.3em}
\subsection{Query expansion}\label{sec:expansion}

Existing query-by-example systems use sketch-based query synthesis approaches~\cite{DBLP:journals/pvldb/TakenouchiIOS21,Wang2017SynthesizingHE, mell2021synthesizing} to enumerate candidate queries. A sketch query is a query with unspecified parts in the form of holes and these approaches enumerate the search space by first generating high-level sketch queries and then filling them with low-level details.
However, enumerative search is slow and memory-intensive. Inspired by execution-guided synthesis approaches~\cite{DBLP:conf/iclr/ChenLS19, DBLP:conf/icml/HuangSBSAN20}, which treat a program as a sequence of manipulations and use the results of partial programs to guide the search, \system explores the search space based on the results of executing intermediate queries
on the examples. Instead of synthesizing sketch queries with uninstantiated holes that cannot be executed directly, or directly applying techniques from~\cite{DBLP:conf/iclr/ChenLS19, DBLP:conf/icml/HuangSBSAN20}, which require a large amount of data to train a neural synthesizer, \system expands queries by adding instantiated and executable constraints, executing partial queries to assess the promise of each explored path, and iteratively refining a query towards the target query.

In our synthesis algorithm, method $\textsc{ExpandQuery}(q, P)$ takes as input a query $q$ to expand, and a set of user-defined functions, $P$ to construct more complex queries. The function returns a set of expanded queries as illustrated in \Cref{fig:algorithmExample}(a).

Our query synthesis approach uses a compact query notation, which can be seen as a DSL. The DSL captures the logical structure of the queries to synthesize and key query parameters, but omits the details of the full, underlying SQL (or Datalog). This approach is important for several reasons: First, since we do not generate arbitrary SQL, but rather queries that conform to the structure presented
in Section~\ref{subsec:query_language}, the DSL captures that structure precisely, simplifying the search space and guiding synthesis toward the correctly-structured queries. Second, this approach helps to decouple the logical query specification from the details of the SQL queries that are ultimately executed. As we present in \Cref{sec:query-execution}, critical optimizations are necesssary during the translation from our DSL to SQL to achieve efficient query execution.

In our DSL, we use a \textit{variable} $o$ to represent an object in a query. Different variables represent objects with different $oid$'s. All predicates of a region graph are connected by commas and are represented with shorthand notations that specify only their key-value pairs (for property attributes, e.g., $\texttt{Color}(o_1, \textrm{`cyan'})$), value (for state attributes, e.g., $\texttt{Bottom}(o_1)$), or class (for objects and relationships, e.g., $\texttt{Car}(o_1)$). Then, region graphs are connected in sequence with semicolons. For example, the query for the event from~\Cref{subsec:datamodel} can be represented as:

{\footnotesize
\setlength{\abovedisplayskip}{0pt}
\setlength{\belowdisplayskip}{0pt}
\setlength{\abovedisplayshortskip}{0pt}
\setlength{\belowdisplayshortskip}{0pt}
\begin{align*}
    \vspace{-2em}
    q=&(\texttt{Car}(o_1), \texttt{Motorcycle}(o_2), \texttt{LeftOf}(o_1,o_2), \texttt{Bottom}(o_1));\\
    &(\texttt{Car}(o_1), \texttt{Motorcycle}(o_2), \texttt{RightOf}(o_1,o_2), \texttt{Bottom}(o_1))
\end{align*}
}%
We further use the notation $\texttt{Duration}(g, d)$ to require that the region graph $g$ exist in \textit{at least} $d$ consecutive frames.

During query synthesis, \system expands queries written in our DSL.
\textsc{ExpandQuery} takes any of the following three \textit{actions}:
(i) Graph construction (GC): Add a predicate to an existing region graph.
(ii) Sequence construction (SC): Insert a new region graph consisting of one predicate into any position of the existing sequence of region graphs.
(iii) Duration refinement (DR): Increment the duration constraint of an existing region graph in the sequence.

As shown in~\Cref{fig:algorithmExample}(a), \system starts with an empty query $q_0=q_\emptyset$. In iteration 1, \system takes action SC to expand $q_0$. This results in $q_1$ to $q_5$, each consisting of a single region graph with one predicate drawn from $P$.
In iteration 2 of \Cref{fig:algorithmExample}(a), \system first expands $q_1=\texttt{LeftOf}(o_1,o_2)$. Performing action GC leads to $q_6=(\texttt{LeftOf}(o_1,o_2)$, $\texttt{Bottom}(o_1))$;  SC leads to $q_7=\texttt{LeftOf}(o_1,o_2); \texttt{Bottom}(o_1)$; and DR leads to $q_8=\texttt{Duration}(\texttt{LeftOf}(o_1,o_2), 5)$, assuming the granularity of \texttt{Duration} is 5 frames.

\exampleExpansionFigure

Deciding which actions to take to expand queries is an important design decision in \system.
Imagine the search space as a directed acyclic graph (DAG), where each node represents a possible query in the search space, the root node represents the empty query $q_\emptyset$, and the query of each child node is constructed by applying one action to its parent node. We consider two extreme expansion rules: A restrictive rule (\Cref{fig:exampleExpansiona}) only allows each query to have one construction path in the DAG, while a relaxed rule allows each query to be generated through all possible permutations of actions with multiple construction paths (\Cref{fig:exampleExpansionb}). Our early experiments showed that the restrictive rule led to poor performance: In some cases, queries with high performance were not found because their ancestor queries (i.e., intermediate parent queries) were not good and were getting pruned. Since different predicates and their combinations have different selectivity and contribute differently to the final query performance, the relaxed rule allows the algorithm to focus on synthesizing the more dominant part of the query during early iterations, avoiding all paths to the target query being accidentally pruned. The cost of a relaxed rule is a larger search space and slower query synthesis. In the example of~\Cref{fig:exampleExpansion}, queries are constructed following the paths highlighted in blue, assuming only the best query is expanded at each step.
The restrictive rule selects the query $B;C;A$ with a score of $0.5$ rather than $A;B;C$ with a score of $0.9$, because $A$ has a lower F1 score than $C$ and $B$. On the other hand, using the relaxed rule selects $A;B;C$ by first constructing $C$, then adding $B$ and $A$ to the sequence.
\system follows the relaxed rule: GC can add any predicates that are not already in the region graph (predicates of the same user-defined function but different variables are considered different); SC can insert a new region graph before and after any existing region graph; DR can increment the duration constraint of any existing region graph.

\vspace{-1em}
\subsection{Beam search}\label{sec:pruning}

When traversing the search space, we can greedily expand only the top query or exhaustively expand all. The former is quick to compute but may not find the target query at the end, while the latter is optimal but comes with a prohibitive computational cost. \system uses beam search to balance query performance and synthesis efficiency. Beam search has a wide range of applications in problems with large search spaces, especially in NLP~\cite{DBLP:conf/naacl/LuWWJKKBQYZSC22, DBLP:journals/ml/KumarVME13, vijayakumar2016diverse}. The greedy approach can be viewed as a special case of the beam search with a beam width of one, and the exhaustive search is equivalent to a beam search with an infinitely large beam width. Beam search reduces runtime by limiting the number of explored branches at each iteration. However, the search outputs are not guaranteed to be optimal. Search quality depends on how branches are expanded, scored, and pruned. In our approach, method $\textsc{SampleQueries}(S', L, bw)$ retains the top $bw$ queries to expand by evaluating the F1 score of the $S'$ candidate queries on the set of labeled video segments $L$ (see \Cref{fig:algorithmExample}(c)).
Prior work has analyzed the theoretical properties of beam search under certain assumptions (e.g., monotonicity of scoring functions~\cite{DBLP:journals/tacl/MeisterCV20}, Bayes optimality of tree models under approximations~\cite{DBLP:conf/icml/ZhuoXDZLXG20}). While in this paper, we directly use the F1 score as the scoring function and to retain top queries, other more sophisticated scoring functions and pruning techniques can be used as alternatives~\cite{DBLP:journals/tacl/MeisterCV20, vijayakumar2016diverse, DBLP:journals/speech/AbdouS04}. We show the effectiveness of our synthesis algorithm empirically in~\Cref{sec:evaluation}.

\subsection{Active learning}\label{sec:active_learning}

One challenge with asking the user for only a handful of examples of the intended event is that we risk overfitting, but asking the user to find a larger number of initial examples is difficult. To address this challenge, we use active learning during query synthesis.

Many active learning methods have been proposed in the literature with the goal of labeling samples that maximally improve a model trained on those samples. E.g., uncertainty-based sampling~\cite{Lewis1994HeterogeneousUS}, estimated error reduction~\cite{Roy2001TowardOA}, and core-set approach~\cite{DBLP:conf/iclr/coreset}. In our work, we use active learning to label samples that help us identify which candidate queries are better than others. To do so, we need to label samples where candidates disagree: Therefore, we use disagreement-based active learning~\cite{DBLP:conf/aistats/KarimiGKR0021}.
At every iteration, \system asks the user for a handful of additional labels to differentiate between the $bw$ queries retained from the previous iteration. Intuitively, this method picks the video segments where the retained queries disagree the most.
A similar idea appears in the query-by-committee algorithm~\cite{Seung1992QueryBC} and has been used in other program synthesis and query synthesis systems~\cite{Ji2020QuestionSF, mell2021synthesizing}. \system's contribution is to incorporate active learning in each iteration of the search process, ensuring an interaction-level labeling experience.
Using active learning reduces the number of data points a user has to provide; additionally, we posit that it is easier for users to label system-selected examples than to come up with their own.

During each call to $\textsc{PickNextSegments}(L, U, S')$, \system computes a score for a sample of unlabeled video segments $U$ over the candidate queries $S'$ and then picks the video segments with the largest disagreement.
The score of each sampled, unlabeled video segment is computed as the weighted disagreement between the candidate queries. The weight of each candidate query is set to its regularized performance over the labeled set $L$ (\Cref{subsec:regularization}).
While the algorithm in~\cite{DBLP:conf/aistats/KarimiGKR0021} is designed for the setting where labeling candidates are streamed, \system instead maintains a pool of candidates. Therefore, for each call to \textsc{PickNextSegments}, \system computes the score for a sample of unlabeled video segments $s_v$ over a sample of candidate queries $s_q$ and then picks the best one.
The function updates $L$ and $U$ given the new user labels. As shown in~\Cref{fig:algorithmExample}(a), \system generates five candidate queries $q_1$ to $q_5$ in iteration 1. Among them, $q_1$ has the highest score 0.6, while $q_2$ and $q_3$ give the same second highest score 0.4. \system requests a new video segment to be labeled (\Cref{fig:algorithmExample}(b)), which distinguishes $q_3$ from $q_2$ (\Cref{fig:algorithmExample}(c)).

Before query synthesis begins, users can set the labeling budget as a hyperparameter. \system precomputes the number of iterations of query expansion it will perform and uniformly divides the labeling budget among iterations.
At every iteration, \system asks for new labels only if it has a labeling budget left to do so; otherwise, it proceeds to synthesize queries without requesting new labels.
\system precomputes the number of iterations by relying on the hyperparameters (\Cref{sec:query-execution}) that bound the search process and the query expansion rules (\Cref{sec:expansion}), which increments the complexity of synthesized queries in each iteration.

\subsection{Retaining top queries}\label{sec:retaining_topk_queries}
At the end of each iteration, \system updates its list of candidate final queries in order to retain $k$ best performing ones, as measured by their F1 scores on the labeled
dataset. $\textsc{RetainTopQueries}(Q_t, L, k)$ evaluates all candidate queries $Q_t$ on labeled set $L$ and returns the top $k$ (\Cref{fig:algorithmExample}(d)).

In~\Cref{fig:algorithmExample}(d), at the end of iteration 1, \system stores $q_1$ in $Q_t$ because $k{=}1$, and $q_1$ has the highest score. When the algorithm terminates, \system returns $q_i$ as the final query. In our evaluation, we set the default value of $k$ to 100 (and to 1000 for a more complex dataset), which ensures both low overhead and high-performing final queries.

\subsection{Regularization}\label{subsec:regularization}
\system retains top queries using their F1 scores on the smaller $L$ labeled set. This can lead to overfitting since there are likely too few training examples to accurately evaluate each candidate query.
To prevent overfitting, we regularize the score of each candidate query by adding the term: $score_{\textrm{reg}}(q)=score(q)-\lambda\cdot R(q)$, where $\lambda$ controls the importance of the regularization term. $R(q)$ represents the complexity of the query $q$: $ R(q) = \sum_{i=1}^k \big( \alpha_1 n_{pi} + \alpha_2 n_{di} + \alpha_3 n_{di}n_{pi} \big)$. Here, $k$ is the number of region graphs in $q$, $n_{pi}$ is the number of predicates in the $i$th region graph $g_i$, and $n_{di}$ denotes the scale of the duration constraint of $g_i$. $\alpha_1$, $\alpha_2$, and $\alpha_3$ are hyperparameters that control the importance of each term.

\section{Query execution}\label{sec:query-execution}

\executionExampleFigure

As discussed in~\Cref{sec:expansion}, we convert queries in our DSL into SQL and use a relational engine (PostgreSQL in our prototype) to execute them.
In this section, we discuss important optimizations that we apply during this translation process.

\noindent \textbf{Query translation algorithm.}
In~\Cref{subsec:query_language}, we show that queries over videos are naturally recursive. Such queries, however, are slow to execute.
To avoid recursion, we leverage the observation that those queries express the idea of iteratively matching a region graph in
a \textit{contiguous sequence} of frames, rather than arbitrary recursion.
Therefore, we can express them using window functions instead of recursion.

As a second optimization, we note that a query $q$ in our DSL (\Cref{sec:expansion})
takes as input a set of video segments $V$ and returns all segments in $V$ containing at least one event that matches $q$. Therefore, instead of finding all satisfying events for each video segment, the query execution needs to find only one. We leverage this observation to reduce intermediate result sizes
by producing SQL that efficiently computes the \textit{earliest} matching sequences for each region graph specification.
This optimization applies to queries without window specifications and whose duration constraints are of the form ``>'' or ``>='', which are
the only types of queries that \system currently generates.

As a concrete example, consider~\Cref{fig:executionExample}. Assume we want to execute a query $q=\texttt{Duration}(g_1, 2); \texttt{Duration}(g_2, 2); g_3$, where $g_1=\texttt{Far}(o_1, o_2)$, $g_2=\texttt{Near}(o_1, o_2)$ and $g_3=\texttt{Behind}(o_1, o_2)$. Also, we only consider one permutation of objects for each video segment in this example for simplicity. The event that $q$ looks for consists of a contiguous sequence of $g_1$ for at least $d_1=2$ frames, followed by a contiguous sequence of $g_2$ for at least $d_2=2$ frames, etc.

Consider video segment $v_1$. For the first region graph $g_1$, we only need to extract the earliest matching segment of $g_1$, which is $e_{11}=\{g_{12},g_{13}\}$. This is because, for every satisfying event for the query, we can replace its matching segment of $g_1$ with the earliest one and the resulting event will still be a match to the query. $e_{2}=\{g_{13},g_{14},g_{25},g_{26},g_{37}\}$ is a matching event for $q$. By replacing $e_{12}=\{g_{13},g_{14}\}$ with $e_{11}=\{g_{12},g_{13}\}$, we get $e_{1}=\{g_{12},g_{13},g_{25},g_{26},g_{37}\}$, which is still a matching event for $q$. For subsequent region graph $g_2$, we first use the earliest matching segment from $g_1$ to retain matching frames of $g_2$ that are temporally after the end frame of the segment, which are $g_{25}$ and $g_{26}$. Then, we can use the same method as before to find the earliest matching segment from $g_2$, which is $e_{21}=\{g_{25},g_{26}\}$. The same procedure can be applied to $g_3$, giving us $e_{31}=\{g_{37}\}$. Thus, the earliest matching event for $q$ in $v_1$ is $e_{1}$. Video segment $v_2$ does not contain any matching event of $q$.

\begin{figure}[t!]
    \centering
    \begin{subfigure}[t]{.33\linewidth}
        \centering
        \begin{tabular}{c}
            \begin{lstlisting}
CREATE VIEW g1_windowed AS (
    SELECT vid, fid,
        oid1, oid2,
    lead(fid, d1 - 1, 0) OVER (
        PARTITION BY vid, oid1,
        oid2 ORDER BY fid
    ) as fid_offest
    FROM g1
);
            \end{lstlisting}
        \end{tabular}
        \caption{Q1: window}\label{fig:sql_a}
    \end{subfigure}%
    \begin{subfigure}[t]{.33\linewidth}
        \centering
        \begin{tabular}{c}
            \begin{lstlisting}
CREATE VIEW g1_iterated AS (
    SELECT vid,
        min(fid_offest) AS fid,
        oid1, oid2
    FROM g1_windowed
    WHERE fid_offest
        = fid + (d1 - 1)
    GROUP BY vid, oid1, oid2
);
            \end{lstlisting}
        \end{tabular}
        \caption{Q2: iterate}\label{fig:sql_b}
    \end{subfigure}
    \begin{subfigure}[t]{.33\linewidth}
        \centering
        \begin{tabular}{c}
            \begin{lstlisting}
CREATE VIEW g2_filtered AS (
    SELECT t1.vid, t2.fid,
        t1.oid1, t1.oid2
    FROM g1_iterated t1, g2 t2
    WHERE t1.vid = t2.vid
        AND t1.oid1 = t2.oid1
        AND t1.oid2 = t2.oid2
        AND t1.fid < t2.fid
);
            \end{lstlisting}
        \end{tabular}
        \caption{Q3: filter}\label{fig:sql_c}
    \end{subfigure}
    \caption{SQL snippets. \system generates efficient SQL queries by avoiding recursions and pruning intermediate results.}\label{fig:sql_snippets}
\end{figure}

The above algorithm can be expressed using SQL queries, as shown in~\Cref{fig:sql_snippets}. Q1 partitions the data on video ID and each unique permutation of groups of objects so that each group of objects is analyzed separately, and sorts the data on frame ID. Next, for each row, it gets the current frame ID as well as the frame ID that is $d_1$ frames after the current frame ID, or 0 if no such frame exists. The result of Q1 gives a table of intervals. Q2 then finds the earliest matching segment for each combination of video and groups of objects. Together, Q1 and Q2 find the earliest matching segment for $\texttt{Duration}(g_1, 2)$ without using recursive joins. Q3 uses the earliest matching segment from $g_1$ to determine where to look for matching segments in $g_2$.
The example SQL snippets only contain two objects \texttt{oid1} and \texttt{oid2}. In general, Q1 and Q2 partition and group the data by all objects in the region graph $g_i$, and Q3 specifies all the constraints on objects across two region graphs in the where clause.
We demonstrate the impact of both optimizations in \Cref{subsec:microbenchmark}.

\noindent \textbf{Caching.}
We implement an application-level cache to reuse sub-query results. \system generates SQL queries as described above to find matching region graphs in the sequence specified by the DSL query one by one. We cache the intermediate results of all prefixes on the set of video segments to allow other queries sharing the same sub-queries to reuse the results. For example, when executing the query $q=g_1; g_2; g_3$ on video segments $v_1$ and $v_2$, we cache the results of $q_1=g_1$, $q_2=g_1; g_2$, and $q_3=g_1; g_2; g_3$ on $v_1$ and $v_2$. Later on, if a query $q'=g_1; g_2; g_4$ is executed on $v_1$ and $v_3$, \system will reuse the cached results of $q_2=g_1; g_2$ on $v_1$ to improve performance.

\noindent \textbf{Bounding the search space.}
\system synthesizes queries up to a certain maximum size to ensure that the synthesis algorithm eventually terminates. Assume the number of user-defined functions for object classes and attributes is $m_1$ and the number of user-defined functions for relationships is $m_2$. We bound the query search space using a set of hyperparameters: the maximum number of region graphs ($n_g$)  in a query, the maximum number of predicates ($n_p$) in the query, the number of possible values a duration constraint can take ($n_d$), and the number of unique objects allowed in a query ($n_v$). The size of the search space is then $|Q| = O((n_g(n_v m_1 + n_v^2 m_2))^{n_p}\cdot n_d^{n_g})$.

\noindent \textbf{Predicate evaluation.}
The current implementation of \system populates the \texttt{Objects} relation and the property attributes in the \texttt{Attributes} relation before query synthesis (see \autoref{table:relational_schema}), by executing ML models on all video frames. The \texttt{Relationships} relation and the state attributes in the \texttt{Attributes} relation are computed lazily during query execution since they do not require ML models in our prototype, and are thus inexpensive to compute. In general, optimizing what information is computed before query synthesis and what is computed lazily is not the focus of this paper.

\trajectoryQueries
\warsawQueries

To simplify SQL query generation, we implement all predicates (except for join predicates) in \system as user-defined functions
that take attributes and bounding box coordinates as input and return boolean values. In our prototype, we use PostgreSQL and implement user-defined functions for state and relationship predicates that operate on bounding boxes, and for property predicates that operate on key-value pairs.

While \system provides the above user-defined functions, the user can also provide additional user-defined functions incrementally as they use the system, and they can be shared and reused across queries and users~\cite{DBLP:conf/acl/WangGLM17}. In practice, \system has to limit the number of user-defined functions to a reasonable amount to ensure the efficiency of query synthesis.

\section{Evaluation}~\label{sec:evaluation}

We conduct an experimental evaluation of \system.
First, we show that on both synthetic and real-world datasets, compared to existing systems, \system reduces the query synthesis time by 1-2 orders of magnitude, achieves comparable or better F1 scores under the same labeling budget, is more robust to noisy data, and explores and executes queries more efficiently.
Second, we show that \system is capable of synthesizing more complex, flexible queries over arbitrary scene graphs, which existing systems cannot handle.
Third, We conduct a user study to demonstrate the effectiveness and usability of \system. We further compare \system with a machine learning method and conduct an ablation study by varying the various design choices outlined in the previous sections.

\noindent\textbf{Baselines.}
To the best of our knowledge, there are no existing systems that can synthesize queries over scene graphs. The most similar system to \system is Quivr~\cite{mell2021synthesizing}, which synthesizes queries over trajectories. Since object trajectories can be represented using scene graphs, we compare our system against Quivr. We also compare against PATSQL~\cite{DBLP:journals/pvldb/TakenouchiIOS21}, which is a state-of-the-art query-by-example system for relational data.

\noindent\textbf{Predicates.}~\label{subsec:predicates}
For the synthetic dataset, we use 13 predicates in experiments: six relationship predicates (\texttt{Near}, \texttt{Far}, \texttt{LeftOf}, \texttt{RightOf}, \texttt{FrontOf}, and \texttt{Behind}), four state predicates (\texttt{Left}, \texttt{Right}, \texttt{Top}, \texttt{Bottom}), and three property predicates (\texttt{Color}, \texttt{Material}, and \texttt{Shape}). We consider eight colors, two materials, and three shapes. Property predicates are only used for the scene graphs dataset. For duration constraints, we consider three possible values: $\textrm{duration}(g) \geq$ 5, 10, or 15 frames. \Cref{sec:query-execution} discussed how \system evaluates these predicates.
For the real dataset, we use nine predicates: two relationship predicates (\texttt{Faster}, \texttt{Near}) and eight state predicates (\texttt{LaneK}, \texttt{Stopped}, \texttt{HighAccel}). \texttt{LaneK} detects whether a car is in lane K, and we identify five lanes in the video.

\noindent\textbf{Data.} We evaluate our system on both synthetic and real-world datasets. We first evaluate on the CLEVRER dataset~\cite{CLEVRER2020ICLR}, which comprises synthetic videos of moving objects. This dataset includes a variety of geometric shapes interacting in space and time, and comes with ground truth data, facilitating testing queries with varying complexities. We create two benchmarks from this data: trajectories and scene graphs datasets.

\mainResult

\noindent\textbf{Trajectories dataset.}
We create a first dataset to test queries over trajectories, which baseline systems support. We extract 10,080 pairs of object trajectories that overlap in time from 500, 5-second video segments.
Each trajectory pair is essentially a temporally ordered sequence of bounding box pairs $b=(x_{a1}, y_{a1}, x_{a2}, y_{a2}, x_{b1}, y_{b1}, x_{b2}, y_{b2})$ representing two objects in a video segment. We manually generate a set of queries with varying complexities (\Cref{table:trajectory_queries}). To generate ground truth labels, we run each target query on the dataset. We sample 500 trajectory pairs as training data (i.e., data used during query synthesis) and use the rest as test data (i.e., to measure the quality of synthesized queries).

\noindent\textbf{Scene graphs dataset.}
This more complex dataset contains scene graphs extracted from the CLEVRER dataset, which baseline systems do not support. We extract scene graphs from 10,000, 5-second video segments. For every frame of a video segment, we store the object track ID ($oid$), bounding box coordinates, and object attributes (shape, color, and material) of every object in the frame (the \texttt{Objects} and \texttt{Attributes} relations in~\Cref{table:relational_schema}).
We automatically generate three classes of queries with different complexities: \textit{easy}, \textit{medium}, and \textit{hard}. Each generated query contains exactly three variables (i.e., three distinct objects).
Easy queries have exactly three predicates on relationships and states, one property predicate, one region graph, and no duration constraints; medium queries have exactly five predicates on relationships and states, two property predicates, three region graphs, and no duration constraints; hard queries have the same complexity as medium queries but also include duration constraints with three possible values.
Each class contains 40 queries.

\noindent\textbf{Real-world dataset.} We test on a 50-minute traffic video from the YTStreams dataset~\cite{DBLP:conf/sigmod/BastaniHBGABCKM20}. The dataset comes with car trajectories, and we further extract the velocity and acceleration of each car trajectory as the object attributes, following the same procedure described in~\cite{mell2021synthesizing}. From the video, We create 72,159 pairs of overlapping car trajectories, using half of them as training data and the rest as test data.
As shown in~\cref{table:warsaw_queries}, we adopt 13 queries from~\cite{mell2021synthesizing} for our evaluation, with some adjustments, including the use of a different in-lane detection method, removal of real-valued parameters from predicates, and expression of queries using scene graphs, and the introduction of additional queries with two scene graphs. These queries capture a wide range of car behaviors.

\noindent\textbf{Metrics.}
We report the F1 score and query synthesis time. We synthesize queries using the training set and report the F1 score of the query (or the median F1 score if there are multiple queries with the same best score on the training set) over the test set. Each experiment is run 20 times for the trajectories and real-world datasets and five times for the scene graphs dataset, and we report the median F1 score and query synthesis time over these runs.

\noindent\textbf{Implementation details.}
We implement our prototype in Python with PostgreSQL as the backend. We conduct all experiments except for the user study on a computing cluster with Intel Xeon Gold 6230R CPUs at 2.10GHz. When measuring runtimes, we request one node with one core and 100GB of RAM.
Unless otherwise specified, we configure \system as follows. For the trajectories dataset and the real-world dataset, we search for queries with up to 5 predicates across up to 3 region graphs. We use beam width $bw=10$, $\lambda=0.01$ for regularization (with $\alpha_1=\alpha_2=1$, and $\alpha_3=0.1$, see \cref{subsec:regularization}), and we set $k=100$, $s_q$=100, and $s_v=100$.
For the scene graphs dataset, we search for queries with up to 7 predicates, 3 region graphs, and 3 objects. Since the dataset is more complex and challenging, we use a smaller $\lambda$ (because the query complexity term, $R(q)$ has a greater absolute value) and a greater $k$. We use beam width $bw=10$, $\lambda=0.001$ for regularization (with $\alpha_1=\alpha_2=1$, and $\alpha_3=0.1$), and we set $k=1000$, $s_q$=100, and $s_v=100$.

The user study is conducted on an AWS EC2 \texttt{c6id.4xlarge} instance with 16 vCPUs and 32GB of RAM. We search for queries with up to 7 predicates, 3 region graphs, and 3 objects. For duration constraints, we consider three possible values, 25, 50, and 75 frames, which translate to 1, 2, and 3 seconds. To improve the interactivity of the system, we use a smaller beam width $bw=5$ and a smaller sample of candidate queries $s_q=25$ for active learning. We set $s_v=100$, $\lambda=0.001$ and $k=100$.

We consider two variants of Quivr. As per the original paper, we limit the number of atomic predicates in Quivr's queries to 5 and the depth of the nested constructs to 3, which leads to a similar search space as \system. When considering queries without duration constraints (e.g., TQ1-TQ9), we omit Kleene star operators from its search space. Otherwise (e.g., TQ1D-TQ3D), we include Kleene star in the query expansion and add one more predicate $\texttt{MinLength}_\theta$, which checks whether the duration of the input is at least $\theta$ frames. Quivr returns all queries that match the examples, so we select the queries with the simplest structure (which is determined by the number of atomic predicates, and, if the former is the same, by the depth) and report their median F1 score.

Because PATSQL requires that \textit{all} user-provided input tables be used in the solution query,  when comparing against PATSQL, we restrict all systems to only the candidate predicates that appear in the target query. Because PATSQL cannot handle large tables efficiently, we also downsample each trajectory by 75\% (we keep one frame out of four) to reduce the size of the input tables. We refer to this configuration as \textit{simplified tasks}.

\subsection{Results against baselines on trajectories}~\label{subsec:trajectories_results}

We evaluate \system against the two baselines on the trajectories dataset and the set of queries in \Cref{table:trajectory_queries}. For TQ1-TQ9, we omit duration constraints from the search space.
We run each method as follows. For each target query, we randomly select 2 positive and 10 negative examples from the training set and use them as the input to the method.
Each method then asks for $b$ additional examples during the search. For PATSQL, since it is not interactive, we simply sample $b$ more examples randomly from the remaining dataset and provide the $12+b$ examples to the system at the beginning. For Quivr and \system, we input the 12 initial examples, and each system actively requests more examples during the search process.

\textbf{\system is faster than baselines and can find high-performant queries even for complex queries.}
\Cref{table:main_result} shows the query synthesis time of each query to achieve at least a 0.9 F1 score.
For simplified tasks, \system outperforms baselines on 6 queries. PATSQL performs the best or close to the best when queries are simple (TQ1, 5, and 6) because these queries include only two predicates and PATSQL terminates once it finds one solution query. Quivr is slower than \system but comparable to it except for TQ9, in which case Quivr fails due to an insufficient F1 score of the synthesized queries. Under the normal setting, \system is significantly more efficient than Quivr and can find high-performance queries in hundreds of seconds even for complex queries with duration constraints, while Quivr cannot synthesize TQ4, TQ7, TQ9, and TQ1D-TQ3D within 4 hours due to the large number of queries that need to be explored in the enumerative search.

\withoutDurationFigure

\textbf{On simplified tasks, \system and QUIVR find queries with similar F1 scores and outperform PATSQL for the same labeling budget.}
\Cref{fig:withoutDuration} shows the F1 score of each system when varying the user labeling budget on the simplified tasks (which all systems can perform). We assign an F1 score of 0 if a system fails to find a query in 1~hour.
Across all labeling budgets tested, as above, PATSQL performs well when the target query is simplest (TQ1, 5, and 6).
For other queries, PATSQL either fails to find a solution query within 1 hour or the solution query has low performance (below a 0.9 F1 score). When the budget is 30, we observe a decrease in F1 scores for TQ3, 7, 8, and 9, because PATSQL has a less constrained search space than \system and does not scale well when the size of input and output tables increases as more examples are provided.
Quivr performs better than \system when no additional examples are requested (budget=12) because it enumerates the entire search space and finds all consistent queries using the initial examples, while \system prunes candidate queries at every iteration to ensure efficient query synthesis. With a larger labeling budget, \system can request more labels during the search process to select better paths to explore, thus it catches up with Quivr and outperforms it when the budget is 30.

\quivrNoisyData
\noisyFigure

\textbf{\system is more robust to noisy data and produces higher quality queries than Quivr.}
We compare the performance of Quivr and \system when the data is noisy. We do not compare against PATSQL since its performance on video queries is already low even with perfect data.
We inject noise into the original dataset by randomly flipping a fraction of the labels. We vary
the false negative rate from 0.1 to 0.5 and set the false positive rate to 0.1 the false negative one.
\Cref{table:quivr_noisy_data} shows the percentage of runs when the system returned \textit{any} queries for different noise rates. We evaluate the systems over 9 queries (TQ1-TQ9) in the simplified setting. For Quivr, we also report the range of the success rate besides the mean value. When the false negative rate is 0.1, Quivr has a 68\% success rate, and this goes down to only 17\% when the false negative rate is 0.5.
In contrast, \system always returns $k$ queries.
To demonstrate that \system also produces higher quality queries, \Cref{fig:noisy} shows the F1 score of \system and Quivr with a labeling budget of 20, under two different noise rates. When Quivr fails to return any queries, we assign an F1 score of 0. \system performs better than Quivr for different queries and different noise rates.

\queryExplored

\textbf{\system is more computationally efficient than Quivr by exploring fewer queries and executing every query faster.}
\Cref{table:query_explored_and_predictions_per_second} measures the number of queries explored and the number of predictions per second to achieve at least a 0.9 F1 score. We define a prediction as evaluating a query on a video segment, so the number of predictions per second reflects the efficiency of query execution. We evaluate over TQ1-TQ9 in the normal setting and report the numbers for queries that Quivr can synthesize.
By limiting the query search to valid sequences of region graphs and by using the beam search strategy, the number of queries explored by \system is as small as $0.41\%$ of Quivr. \system also executes queries faster than Quivr by up to $5.5\times$ in terms of the number of predictions per second. These two factors together make \system more efficient than Quivr.

\sceneGraphQueriesFigure
\varycpuFigure
\subsection{Results on scene graphs}~\label{subsec:scene_graphs}

Next, we evaluate \system on the scene graphs dataset. Neither PATSQL nor Quivr support such flexible queries at video scale.
Because the dataset is more complex, \system requires that the user provide a slightly larger (although still quite small) number of initial examples to avoid overfitting. We randomly select 15 positive and 15 negative examples from the training dataset as the initial examples.
Later in~\Cref{subsec:microbenchmark}, we discuss how varying the initial number of examples affects the system performance.

\Cref{fig:sceneGraphQueries} shows the F1 score of \system under different user labeling budgets. With a larger budget, \system can learn queries with higher F1 scores, and achieves a median F1 score of 1.0, 0.91, and 0.67 for easy, medium, and hard queries when the budget is 100. Results for a budget of 50 are nearly identical, which shows that \textbf{\system can synthesize high-quality queries within a reasonable budget, even for complex queries}.

\system does not learn good queries in all cases. We note that some queries are easier to synthesize than others, which results in the high variance of the F1 score in \Cref{fig:sceneGraphQueries}. In particular, \system assumes that the ancestor queries of the target query are informative and leverages the performance of those queries to guide the search. However, if the ancestor queries cannot be distinguished from other queries in the same iteration (because they all have the same, typically low, F1 score), \system struggles with learning good queries. One direction of future work could explore other types of more fine-grained user feedback besides binary labels to help \system learn intermediate queries~\cite{DBLP:conf/cvpr/RussakovskyLL15}.

\system's query synthesis time on the scene graphs dataset is greater than on the trajectories dataset. It increases from tens of seconds (\Cref{table:main_result}) to minutes. \textbf{However, \system can easily be parallelized to reduce synthesis time, as executing queries in PostgreSQL is embarrassingly parallel}.
\Cref{fig:varycpu} shows the query synthesis time for the hard, scene-graphs queries when varying the number of cores and for a labeling budget of 50. Query synthesis time decreases significantly by using multiple cores and reduces to 13 minutes with eight cores. \system achieves $1.76\times$ speedup with 2 cores and $3.03\times$ speedup with 8 cores. Note that this is the largest labeling budget for our experiments. With a lower budget, \system is even faster. The query synthesis time can be further reduced by using more cores, further decreasing the labeling budget, decreasing $bw$, or decreasing $k$.

\subsection{Results on real-world dataset}\label{subsec:warsaw}

\warsawFullFigure
\warsawSuccessRate

\userStudyQueries

We compare \system and Quivr on the real-world dataset and queries in \Cref{table:warsaw_queries}. For WQ1-WQ9, we omit duration constraints from the search space. We randomly select 2 positive and 10 negative examples as initial examples. Since the target events are extremely rare in the dataset, we consider \system with two different $s_v$ values, 100 and 500, to increase the number of sampled positive examples during active learning.

\textbf{\system outperforms Quivr in terms of F1 scores and query synthesis time on the real-world dataset.} \Cref{fig:warsaw_full} shows the F1 score and query synthesis time of both systems under different labeling budgets. The timeout is set to 4 hours for both systems. \system can synthesize all queries in hundreds of seconds with a labeling budget of 30, while Quivr suffers from a low success rate (as shown in~\Cref{table:warsaw_success_rate}) and, even when it succeeds, takes thousands of seconds to synthesize queries. \system performs worse or similar to Quivr with a labeling budget of 12, but with some additional examples, \system can easily achieve higher F1 scores. In addition, sampling more candidate video segments during active learning ($s_v$=500) increases the F1 score in most cases but also slightly increases the synthesis time.

\subsection{User study}\label{subsec:user_study}

\userStudyFigure

We conduct a user study to understand the effectiveness and usability of \system. We use internal validation since asking participants to provide their own queries would make it difficult to compare across participants. Our goals include observing the system performance in the presence of user noise, measure task completion time, and collect qualitative feedback. We recruited 18 university students studying Computer Science. The study is conducted over the scene graphs dataset with six query tasks with different complexities, as shown in~\Cref{table:user_study_queries}.

We started each session by explaining our definitions of predicates and walking through an example task with participants to get them familiar with the interface and the task. Each participant then completed three query tasks in sequence (either UQ1, 3, 5 or UQ2, 4, 6) with increasing complexity from easy, to medium, to hard. Each query task described the target complex event using natural language. In the user study, we provided the initial examples with ground-truth labels to \system so that participants could focus on interacting with the system during query synthesis. We randomly selected 10 positive and 10 negative examples for UQ1, 2 and 4 as the initial examples, and 15 for other queries to account for the increased query complexity. \system iteratively selected video segments to label, with a budget of 50 per task. After each task, participants reviewed the synthesized queries and provided subjective ratings and qualitative feedback via a questionnaire.

\textbf{Participants are able to complete the task in a reasonable amount of time.} \Cref{fig:user_study_time} shows the task completion time of each query. The median completion time of each task ranges between 8.6 and 12.3 minutes. The average time to label a video segment for each query task is 22, 16, 25, 20, 30, and 22 seconds, respectively. Also, the system time only takes a small portion of the total time, suggesting that \system is efficient.

\userStudyErrorRate

\textbf{\system is resilient to user noise, whereas Quivr is not.} \Cref{fig:user_study_f1} shows the F1 score of synthesized queries under different labeling scenarios: using the ground-truth labels (\texttt{Perfect}), using real user labels (\texttt{User}), using injected label noise with a false negative rate of 0.3 and a false positive rate of 0.03 (\texttt{Fixed}), and using injected label noise with the same noise rate as users for each query task (\texttt{Simulated}).
Participants' labeling error rates are shown in~\Cref{table:user_study_error_rate}.
Each experiment is run nine times. \textbf{F1 scores of \texttt{User} lie between \texttt{Perfect} and \texttt{Simulated}}. Due to the user noise, the query performance decreases compared to using perfect labels. Because the same error rate is used, \texttt{Simulated}'s performance is worse or similar to \texttt{User}'s, which justifies our expectations for the simulation experiment. \textbf{For queries that \system can synthesize with high quality, participants can help to find them with at least 0.8 F1 scores.} Not surprisingly, for queries (UQ5 \& 6) that \system would fail even with perfect labels, participants cannot instruct the system to achieve better performance using the current prototype.
Interestingly, most video segments that \system selects for hard queries are negative examples, making it difficult to learn good queries. However, these segments are not equally negative, as many participants consider some negative samples to be almost positive that fail to satisfy one predicate. This suggests that \system could potentially benefit from more fine-grained user feedback.

\textbf{Participants would like \system to be more responsive.} In the questionnaire, the average rating of system responsiveness is 3.53 out of 5. Although the average system time across all tasks is only 195 seconds, participants observe greater latency in early and middle iterations, since there are more branches to expand and more candidate queries to evaluate. Taking UQ1 as an example, the average wait time to receive the first video segment in the 5th iteration is 31 seconds, while in the 10th iteration, it is only 1.3 seconds. Other techniques can be used to further hide the latency of the system~\cite{daum2023vocalexplore}, which is not the focus of this work.

\subsection{Ablation study}~\label{subsec:microbenchmark}

We study the impact of hyperparameters and the sensitivity of \system to hyperparameter tuning. Unless otherwise specified, we evaluate \system on the scene graphs dataset and use the default values for hyperparameters that we are not studying.

\varyInitExamplesFigure
\heatmapFigure
\textbf{\system synthesizes useful queries with a small number of initial examples, but providing more initial examples improves performance.} \Cref{fig:varyInitExamples} shows \system's F1 score for different numbers of initial examples, from five positives and five negatives (ten in total), to 25 positives and 25 negatives (50 in total), while the total labeling budget remains 100. When the number of initial examples is smallest, \system can incorrectly retain queries that overfit the examples in early iterations, leading to final results with low F1 scores. Results, however, quickly improve, and with as few as 30 initial examples, median F1 scores are high at 0.998, 0.911, and 0.670 for the easy, medium, and hard queries respectively. \Cref{fig:heatmap} shows the F1 score of \system under different numbers of initial examples and queries with different complexity, while the total labeling budget remains 100. As expected, \system obtains higher F1 scores with more initial examples and when the target queries are simpler. The heatmap aggregates the results of all easy, medium, and hard queries, and we use $\alpha_1=1,\alpha_2=0.9,\alpha_3=0.2$ when computing the query complexity for better visualization. Because manually finding initial examples requires more user effort than labeling video segments selected by \system, our approach is to balance user effort and system's performance, using 30 initial examples in all other experiments on the scene graphs dataset. Interestingly, since \system can produce good queries in many cases even with a few examples, users always have the option to first start with a small set and restart with more examples if needed.

\activeLearningFigure
\textbf{Active learning helps \system learn better queries quicker than randomly sampling video segments to label.}
\Cref{fig:activeLearning} shows \system's F1 score when either using active learning or randomly selecting additional examples during query synthesis. We set the labeling budget to 50, and we vary the number of initial examples.
The improvement of active learning over random sampling is more significant when the number of initial examples is larger since fewer labels are requested per iteration and the selection of informative examples becomes more important.

\varybwFigure
\varylambdaFigure
\textbf{Increasing the beam width improves performance, but also increases the synthesis time}. \Cref{fig:varybw} shows the F1 score of \system under different values of $bw$.
Increasing $bw$ from 1 to 10 increases median F1 scores on easy queries by $3.10\%$, on medium queries by $23.8\%$, and on hard queries by $34.3\%$, but increasing from 10 to 20 only further improves median scores by up to 3.07\%. At the same time, increasing $bw$ from 10 to 15 increases the median synthesis time on hard queries by $27.8\%$. \system defaults to $bw=10$ to strike a balance between query performance and synthesis time.

\textbf{Regularization helps \system learn better queries in many cases and does not harm performance in other cases.} \Cref{fig:vary_lambda} shows the F1 score of \system on both the trajectories dataset and the scene graphs dataset, under different values of $\lambda$. For the trajectory pairs dataset, we set the labeling budget to 20 and the initial number of examples to 12 (i.e., two positive and ten negative examples); for the scene graphs dataset, we set the labeling budget to 50 and the initial number of examples to 30. For the trajectories dataset, we see an improvement for many queries with regularization, especially when the queries are simple (e.g., with fewer predicates). With $\lambda=0.01$, we see a median improvement between $44.6\%$ and $-3.44\%$ for the queries from~\Cref{table:trajectory_queries}. TQ4, TQ9, TQ2D, and TQ3D are more complex than other queries and have four predicates, so applying regularization is unlikely to improve the performance. For the more complex scene graphs dataset, regularization does not improve F1 scores but also does not compromise system performance. With $\lambda=0.001$, the median change in F1 score stays between $2.84\%$ and $-7.69\%$. For this reason, \system defaults to using a small regularization factor.

\varykFigure
\textbf{increasing $k$ slightly improves performance.} We evaluate the F1 score of \system under different values of $k$, as shown in~\Cref{fig:varyk}. For medium and hard scene graph queries, increasing $k$ from $1$ to $1000$ slightly improves the median F1 score by $2.36\%$ and $9.12\%$ respectively. Since the target queries are complex, they are not explored until later iterations of the algorithm, when most labeling budget has been used. Labels requested afterward have little impact on the ranking of the queries in the list, so the performance improvement from increasing $k$ is also limited. Easy queries already have near-perfect F1 scores, so increasing $k$ does not improve performance.

\queryExecutionFigure
\textbf{\system's query execution optimizations significantly reduce query synthesis times.}
\Cref{fig:query_execution} shows the average \textit{query execution time} for variants of our query translation algorithm on scene graph queries. We generate SQL queries using our complete query translation algorithm, and two variants that omit optimizations of pruning intermediate results and avoiding recursion, respectively, and then execute them over all 10,000 video segments from the scene graphs dataset in PostgreSQL. We evaluate \system on SQ1-SQ3, and each query is executed 20 times. As the figure shows, the query execution time increases dramatically without either optimization. Pruning intermediate results boosts performance by up to $1.66\times$. Recursion avoidance is important when the query contains duration constraints (SQ1 and SQ3), providing up to a $4.14\times$ speedup.

\Cref{fig:caching} shows the impact of our caching mechanism. Caching query results leads to a $1.97\times$ speedup. We experimented on hard queries using four CPUs and a labeling budget of 50.

\subsection{Comparing with machine learning method}\label{subsec:ml_methods}

\machineLearningMethods

To determine whether a video segment contains an event of interest, \system's query synthesis approach could be substituted by training an ML model.
However, this approach requires a large number of labeled examples and complicates interpretability~\cite{DBLP:journals/pvldb/FarihaM19}. We compare \system against an existing ML approach that builds domain-specific models for videos~\cite{daum2023vocalexplore} on the scene graphs dataset. For \system, we adopt the same setting as outlined in~\Cref{subsec:scene_graphs}, with a labeling budget of 50. The ML approach extracts features using pretrained video and image models and trains a linear model using the same set of user-labeled video segments selected by \system. We use both \textsc{MViT}~\cite{DBLP:conf/iccv/0001XMLYMF21} and \textsc{CLIP}~\cite{DBLP:conf/icml/RadfordKHRGASAM21} as the feature extractors. \Cref{table:machine_learning_methods} shows that \textbf{\system achieves higher F1 scores than the ML approach for all queries, even when using \textsc{CLIP} with all 500 training samples.}
\section{Related work}\label{sec:related_work}

\textbf{Video analytics systems.}
Many recent VDBMSs have been proposed and focus on a wide range of data management challenges, including fast inference over videos~\cite{DBLP:journals/pvldb/KangBZ19, DBLP:conf/sigmod/LuCKC18, Moll2020ExSampleES, DBLP:conf/sigmod/BastaniHBGABCKM20, DBLP:conf/icde/AndersonCRW19}, storage optimization~\cite{DBLP:conf/icde/tasm, DBLP:conf/sigmod/vss, DBLP:conf/eurosys/vstore}, efficient dataflow processing~\cite{10.1145/3197517.3201394}, preprocessing and indexing~\cite{DBLP:conf/sigmod/BastaniM22, DBLP:conf/sigmod/KangGBHZ22, DBLP:conf/sigmod/HeASC20}, exploration and organization~\cite{DBLP:conf/cidr/DaumZHBHKCW22}, privacy~\cite{DBLP:conf/nsdi/CangialosiAANSN22}, and tuning configurations~\cite{DBLP:conf/cloud/RomeroZYK21, DBLP:conf/sigcomm/JiangABSS18, DBLP:conf/sigmod/HeC22}. Those techniques are orthogonal to our work.

\textbf{Compositional query processing over videos.}
Several systems have explored compositional queries over videos~\cite{DBLP:conf/sigmod/Chen0K20, DBLP:conf/sigmod/Chen0KY21, DBLP:journals/pvldb/BastaniMM20, DBLP:conf/sigmod/ChaoKX20, DBLP:journals/pvldb/ChenKYY22, DBLP:journals/corr/abs-1910-02993, DBLP:conf/sensys/LiuGUMCG19, DBLP:journals/corr/abs-2104-06142, DBLP:conf/bigdataconf/YadavC19, DBLP:journals/tmm/KuoC00, mell2021synthesizing}. However, they either require that users express compositional queries explicitly~\cite{DBLP:journals/pvldb/ChenKYY22, DBLP:journals/corr/abs-1910-02993, DBLP:conf/sensys/LiuGUMCG19}, or train customized models for such events~\cite{DBLP:journals/corr/abs-2104-06142, DBLP:journals/pvldb/BastaniMM20, DBLP:conf/sigmod/ChaoKX20}. Instead, \system uses a query-by-example approach to minimize user effort and learn and refine a query specification iteratively from user feedback.

\textbf{Accelerating query execution over visual data.}
Prior work has focused on accelerating queries by pre-filtering frames to avoid expensive computation~\cite{DBLP:conf/sigmod/LuCKC18, DBLP:journals/pvldb/HeDCB21, he2023masksearch}, optimizing the sampling rate~\cite{Moll2020ExSampleES, DBLP:conf/sigmod/BastaniHBGABCKM20}, and building specialized models~\cite{DBLP:journals/pvldb/KangBZ19, DBLP:conf/icde/AndersonCRW19}. \system currently extracts all objects in a video before the search process, but could leverage those existing techniques to avoid running expensive object detection and tracking algorithms on all frames.

\textbf{Query by example.}
Many systems have been proposed for SQL queries over relational data~\cite{DBLP:conf/sigmod/PsallidasDCC15, DBLP:journals/pvldb/FarihaM19, DBLP:journals/pvldb/TakenouchiIOS21, DBLP:journals/pvldb/LiCM15, Wang2017SynthesizingHE}. \system focuses on learning queries for video events, with a more constrained form than general relational queries. Quivr~\cite{mell2021synthesizing} is most similar to our work, but it only operates over object trajectories and assumes noiseless inputs. SQuID~\cite{DBLP:journals/pvldb/FarihaM19} synthesizes queries based on semantic similarity but does not support self-joins and non-equi joins, which are necessary to find our video events, and S4~\cite{DBLP:conf/sigmod/PsallidasDCC15} ranks queries based on input containment and is limited to project-join queries.

\textbf{Program synthesis.}
Program synthesis has been used for a wide range of tasks, such as generating referring relational programs~\cite{DBLP:conf/icml/HuangSBSAN20}, authoring visualizations~\cite{DBLP:conf/chi/WangFBDCK21,DBLP:conf/chi/PangRJ22}, learning relational data transformation~\cite{DBLP:conf/chi/KandelPHH11}, and synthesizing programs for string processing~\cite{DBLP:conf/iclr/OdenaSBSSD21}. Unlike these systems, \system synthesizes queries in the video domain, which is significantly different in terms of the spatio-temporal complexity and the prevalence of noise.

\textbf{Active learning.} Model Picker~\cite{DBLP:conf/aistats/KarimiGKR0021} uses active learning to distinguish the best model from a set of pretrained classifiers. \system adapts this approach to select the most promising queries to explore. Quivr~\cite{mell2021synthesizing} also prunes candidate queries using active learning, but only after enumerating all candidates. In contrast, \system integrates active learning and labeling into the search process to ensure both synthesis efficiency and query performance.
\section{Conclusion}

In this paper, we presented \system, a new system that synthesizes compositional queries from examples. \system models compositional events as spatio-temporal scene graphs, explores the query search space using results of executing intermediate queries and beam search, leverages active learning to reduce user effort, and generates efficient SQL queries to reduce computational effort.
\begin{acks}
This work was supported in part by the NSF through awards CCF-1703051
and IIS-2211133 as well as a grant from CISCO.
\end{acks}

\end{sloppypar}

\bibliographystyle{ACM-Reference-Format}
\bibliography{self}

\end{document}